\documentclass[backref, breaklinks, twocolumn, colorlinks]{aastex631}
\hypersetup{linkcolor=blue,citecolor=blue,filecolor=cyan,urlcolor=black}

\usepackage{latexsym,amsmath,amssymb}
\usepackage{verbatim}
\usepackage{color}
\usepackage[hang]{footmisc}
\usepackage{soul,xcolor}

\bibpunct{(}{)}{;}{a}{}{,}

\newcommand {\chandra} {\textsl{Chandra}}
\newcommand {\swift} {\textsl{Swift}}

\newcommand {\nicer} {\textsl{NICER}}

\def \rsun {\ifmmode$R$_{\odot}\else R$_{\odot}$}

\def \hcm {\hbox {\ifmmode $ atoms cm$^{-2}\else atoms cm$^{-2}$\fi}}

\def\approxgt{\mathrel{\hbox{\rlap{\lower.55ex \hbox {$\sim$}}
        \kern-.3em \raise.4ex \hbox{$>$}}}}
\def\approxlt{\mathrel{\hbox{\rlap{\lower.55ex \hbox {$\sim$}}
        \kern-.3em \raise.4ex \hbox{$<$}}}}

\def \arcmin {\hbox{$^\prime$}}
\def \arcsec {\hbox{$^{\prime\prime}$}}

\newcommand {\fluxcgs} {erg~cm$^{-2}$~s$^{-1}$}

\newcommand\T{\rule{0pt}{2.6ex}}
\newcommand\B{\rule[-1.2ex]{0pt}{0pt}}

\def \src {SGR 1830$-$0645}

\begin{document}
\setstcolor{red}

\title{\rm \uppercase{X-ray burst and persistent emission properties of the magnetar\\ \src\ in outburst}}. 

%
\author[0000-0002-7991-028X]{George~Younes}
\affiliation{Department of Physics, The George Washington University, Washington, DC 20052, USA, gyounes@gwu.edu}
\affiliation{Astronomy, Physics and Statistics Institute of Sciences (APSIS), The George Washington University, Washington, DC 20052, USA}

\author[0000-0001-8551-2002]{Chin-Ping Hu}
\affiliation{Department of Physics, National Changhua University of Education, Changhua 50007, Taiwan}

\author[0000-0002-7418-7862]{Karishma Bansal}
\affiliation{Jet Propulsion Laboratory, California Institute of Technology, Pasadena, CA 91109, USA}

\author[0000-0002-5297-5278]{Paul S. Ray}
\affiliation{Space Science Division, U.S. Naval Research Laboratory, Washington, DC 20375, USA}

\author[0000-0002-8912-0732]{Aaron~B.~Pearlman}
\altaffiliation{McGill Space Institute~(MSI) Fellow.}
\affiliation{Department of Physics, McGill University, 3600 rue University, Montréal, QC H3A 2T8, Canada}
\altaffiliation{FRQNT Postdoctoral Fellow.}
\affiliation{McGill Space Institute, McGill University, 3550 rue University, Montréal, QC H3A 2A7, Canada}
\affiliation{Division of Physics, Mathematics, and Astronomy, California Institute of Technology, Pasadena, CA 91125, USA}

\author[0000-0001-6664-8668]{Franz Kirsten}
\affiliation{Department of Space, Earth and Environment, Chalmers University of Technology, Onsala Space Observatory, 439 92, Onsala, Sweden}

\author[0000-0002-9249-0515]{Zorawar~Wadiasingh}
\affiliation{Department of Astronomy, University of Maryland, College Park, Maryland 20742, USA}
\affiliation{Astrophysics Science Division, NASA Goddard Space Flight Center, Greenbelt, MD 20771, USA.}
\affiliation{Center for Research and Exploration in Space Science and Technology, NASA/GSFC, Greenbelt, Maryland 20771, USA}

\author[0000-0002-5274-6790]{Ersin G\"o\u{g}\"u\c{s}}
\affiliation{Sabanc\i~University, Faculty of Engineering and Natural Sciences, \.Istanbul 34956 Turkey}

\author[0000-0003-4433-1365]{Matthew~G.~Baring}
\affiliation{Department of Physics and Astronomy - MS 108, Rice University, 6100 Main Street, Houston, Texas 77251-1892, USA}

\author[0000-0003-1244-3100]{Teruaki Enoto}
\affiliation{Extreme Natural Phenomena RIKEN Hakubi Research Team, Cluster for Pioneering Research, RIKEN, 2-1 Hirosawa, Wako, Saitama 351-0198, Japan}

\author{Zaven Arzoumanian}
\affiliation{Astrophysics Science Division, NASA Goddard Space Flight Center, Greenbelt, MD 20771}

\author{Keith C. Gendreau}
\affiliation{Astrophysics Science Division, NASA Goddard Space Flight Center, Greenbelt, MD 20771}

\author[0000-0003-1443-593X]{Chryssa~Kouveliotou}
\affiliation{Department of Physics, The George Washington University, Washington, DC 20052, USA, gyounes@gwu.edu}
\affiliation{Astronomy, Physics and Statistics Institute of Sciences (APSIS), The George Washington University, Washington, DC 20052, USA}

\author[0000-0002-3531-9842]{Tolga G\"uver}
\affiliation{Istanbul University, Science Faculty, Department of Astronomy and Space Sciences, Beyaz\i t, 34119, Istanbul, Turkey}
\affiliation{Istanbul University Observatory Research and Application Center, Istanbul University 34119, Istanbul Turkey}

\author{Alice~K.~Harding}
\affiliation{Theoretical Division, Los Alamos National Laboratory, Los Alamos, NM 87545, USA}

\author[0000-0002-4694-4221]{Walid A. Majid}
\affiliation{Jet Propulsion Laboratory, California Institute of Technology, Pasadena, CA 91109, USA}
\affiliation{Division of Physics, Mathematics, and Astronomy, California Institute of Technology, Pasadena, CA 91125, USA}

\author{Harsha Blumer}
\affiliation{Department of Physics and Astronomy, West Virginia University, Morgantown, WV 26506, USA}
\affiliation{Center for Gravitational Waves and Cosmology, West Virginia University, Chestnut Ridge Research Building, Morgantown, WV 26505, USA}

\author{Jason W. T. Hessels}
\affiliation{Anton Pannekoek Institute for Astronomy, University of Amsterdam, Science Park 904, 1098 XH, Amsterdam, The Netherlands, ASTRON, Netherlands Institute for Radio Astronomy, Oude Hoogeveensedijk 4, 7991 PD Dwingeloo, The Netherlands}

\author[0000-0003-4056-4903]{Marcin P. Gawro\'nski}
\affiliation{Institute of Astronomy, Faculty of Physics, Astronomy and Informatics, Nicolaus Copernicus University, Grudziadzka 5, 87-100 Toru\'n, Poland}

\author[0000-0003-3655-2280]{Vladislavs Bezrukovs}
\affiliation{Engineering Research Institute Ventspils International Radio Astronomy Centre, Ventspils University of Applied Sciences, Inzenieru street 101, Ventspils,  LV-3601, Latvia}

\author{Arturs Orbidans}
\affiliation{Engineering Research Institute Ventspils International Radio Astronomy Centre, Ventspils University of Applied Sciences, Inzenieru street 101, Ventspils,  LV-3601, Latvia}

\begin{abstract}

We report on NICER X-ray monitoring of the magnetar \src\ covering 223
days following its October 2020 outburst, as well as \chandra\ and
radio observations. We present the most accurate spin ephemerides of
the source so far: $\nu=0.096008680(2)$~Hz,
  $\dot{\nu}=-6.2(1)\times10^{-14}$~Hz~s$^{-1}$, and a significant
second and third frequency derivative terms indicative of
non-negligible timing noise. The phase-averaged 0.8--7~keV spectrum is
well fit with a double-blackbody (BB) model throughout the campaign.
The BB temperatures remain constant at 0.46 and 1.2~keV. The areas and
flux of each component decreased by a factor of 6, initially through a
steep decay trend lasting about 46~days followed by a shallow
long-term one. The pulse shape in the same energy range is initially
complex, exhibiting three distinct peaks, yet with clear continuous
evolution throughout the outburst towards a simpler, single-pulse
shape. The rms pulsed fraction is high and increases from about $40\%$ to $50\%$. We
find no dependence of pulse shape or fraction on energy. These results
suggest that multiple hotspots, possibly possessing temperature
gradients, emerged at outburst-onset, and shrank as the outburst
decayed. We detect 84 faint bursts with \nicer, having a strong
preference for occurring close to the surface emission pulse maximum;
the first time this phenomenon is detected in such a large burst
sample. This likely implies a very low altitude for the burst emission
region, and a triggering mechanism connected to the surface active
zone. Finally, our radio observations at several epochs and multiple
frequencies reveal no evidence of pulsed or burst-like radio
emission.

\end{abstract}

\section{Introduction}
\label{Intro}

The variable emission from magnetars, spanning timescales from
milliseconds to years, constitutes distinctive character that
separates them from the rest of the isolated neutron star
family. Magnetars emit bright, hard X-ray short bursts with
luminosities ranging from $10^{37}$ to $10^{41}$~erg~s$^{-1}$
\citep[e.g.,][]{collazzi15ApJS} and sub-second duration. Often
accompanying these bursts, magnetars enter a period of elevated
persistent flux level which could exceed their quiescent emission by
as many as three orders of magnitude, i.e., outbursts
\citep[e.g.,][]{cotizelati18MNRAS}. Historically, these properties
were observed in magnetars with a narrow range of spin periods around
the $2-12$~seconds and large spin-down rates
$\dot{P}\sim10^{-11}$--$10^{-13}$~s~s$^{-1}$. Assuming magnetic-dipole
braking, these timing properties imply large dipolar field strengths,
typically in the range of $B\sim10^{13}-10^{15}$~G at the
equator, and young spin-down ages of few thousand years
\citep[e.g.,][see also
\citealt{mereghetti08AARv:magentars,kaspi17:magnetars} for
reviews]{kouveliotou98Nat:1806}. The large available magnetic
energy is putatively the main source of power in magnetars since, for
most, the rotational energy loss is incapable of powering their X-ray
output in quiescence and/or during outbursts
\citep{thompson96ApJ:magnetar,thompson02ApJ:magnetars,turolla15:mag}.

While the aforementioned temporal characteristics are certainly
representative of the bulk of the magnetar population, magnetar-like
activity has now been observed from several other classes of isolated
neutron stars: low B-field magnetars, e.g., SGR 0418+5729
\citep[$B\sim 6\times10^{12}$~G,][]{rea13ApJ:0418}, high B-field
rotationally-powered pulsars \citep[PSR J1846$-$0258 and PSR
J1119$-$6127][]{gavriil2008Sci,archibald16:j1119,gogus16:j1119}, and a
central compact object with a very long spin-period of about 6.67~hrs
\citep{rea16:rcw103}. Moreover, a wind nebula, an otherwise common
characteristics of young rotation-powered pulsars, has been detected
around a typical magnetar, Swift~J1834.9$-$0846
\citep{younes16ApJ:1834}. These results strengthen the case for a
connection, perhaps evolutionary, among all isolated neutron stars
\citep{vigano13MNRAS}.

On 2020 October 10, the BAT instrument onboard the Neil Gehrels
  Swift Observatory (\swift) triggered on a short, soft gamma-ray
burst from the direction of the Galactic plane, and prompt XRT
  observations revealed a bright, previously unknown
X-ray source, now dubbed \src\ \citep{page20ATel14083}. Using the same
data set, \citet{gogus20ATel14085} reported the detection of coherent
pulsations with a spin-period of 10.4~s, confirmed with \nicer\ few
hours later \citep{younes20ATel14086}. \nicer\ performed intense
monitoring of the source in the ensuing days, and subsequently (11
days later) revealed the detection of a spin-down rate,
$\dot\nu\approx9.0\times10^{-14}$~Hz~s$^{-1}$, flux decay, and the
detection of numerous short X-ray bursts \citep{ray20ATel14112}. All
of these characteristics fall within the realm of typical magnetar
properties, and implies that \src\ is a new addition to the magnetar
family \citep[see also][]{cotizelati21ApJ1830}. So
far, the source has not been detected in radio at several frequencies
\citep{surnis20ATel14091,maan20ATel14098, cotizelati21ApJ1830}.

In this paper, we present the detailed soft X-ray and radio properties
of \src\ following its first-detected outburst utilizing the \nicer\
X-ray telescope, \chandra\ telescope, and several radio facilities. We
introduce our observations and data reductions in
Section~\ref{obs}. Our results are presented in Section~\ref{res}. We
discuss the implications of our findings in Section~\ref{discuss},
including our assumed fiducial distance of 4~kpc to the source.

\section{Observations and data reduction}
\label{obs}

\subsection{X-rays}

We observed \src\ with \nicer\ starting approximately 4 hours after
its discovery with BAT on 2020 October 10 at 14:29 UTC
\citep{page20ATel14083}. Prior to the sun-constrained period which
started on 2020 November 17, thirty-seven observations were carried
out (obs IDs 3201810101 to 3201810137). We restarted the \nicer\
campaign on 2021 February 10. Our current analysis includes data
extending up to 2021 May 21 (up to obs. ID 4201810118). \nicer\
consists of 56 co-aligned X-ray concentrating optics covering a
30$\arcmin^2$ field of view and providing a collecting area of
1900~cm$^2$ at 1.5~keV. It offers high precision timing and spectral
capabilities, while affording a low background
\citep{gendreau16SPIE}. We processed NICER data using
\texttt{NICERDAS} version v007a, as part of HEASOFT version
6.27.2. Good time intervals were created from raw level 1 event files
using the nicer tool \texttt{nicerl2}. We use the default filtering
criteria as described  in \nicer\ Data Analysis
Guide\footnote{\href{https://heasarc.gsfc.nasa.gov/docs/nicer/data\_a
    alysis/nicer\_analysis\_guide.html}{https://heasarc.gsfc.nasa.gov/docs/nicer/data\_a
    alysis/nicer\_analysis\_guide.html}}, except the underonly\_range
has been relaxed to $0-300$. This background is mainly due to
  optical loading and only affects the low energy part of the
  spectrum, below the 0.8--7~keV range we consider for all of our
  analyses. This choice for the energy range is driven by the hydrogen
  column density in the direction of the source and its soft
  spectrum. We extracted background spectra utilizing the {\texttt
  nibackgen3C50} method \citep{remillard2021:3c50} and added a
conservative 20\% systematic uncertainty to the estimated background
number counts per \nicer-energy channel.

We observed the field of \src\ with the \chandra\ X-Ray Observatory
Advanced CCD Imaging Spectrometer \citep[ACIS,][]{garmire03SPIE} for
4.6 ks on 2020 October 13 (Observation ID: 24841). The target was
positioned at the nominal aimpoint of the S3 chip of the ACIS-S
detectors. The observation was operated in 1/8th subarray mode, which
provided a temporal resolution of 0.4 s. We reduced the data using the
\chandra\ Interactive Analysis of Observations
(CIAO\footnote{https://cxc.cfa.harvard.edu/ciao/} version 4.12) and
the calibration database of Chandra CALDB version 4.9.2.1. 

We utilize Xspec v12.11.0 (Arnaud 1996) for the \nicer\ X-ray spectral
analysis. We do not attempt any spectral analysis of the \chandra\
data due to heavy pile-up. We use the latest \nicer\ response and
ancillary files in the CALDB release \texttt{xti20200722}. We group
the data to have 50 counts per energy bin and utilize the $\chi^2$
statistics for model parameter estimation and error calculation. We
quote all errors at the $1\sigma$ ($68\%$ uncertainty level) unless
otherwise noted. 

\subsection{Radio Campaigns}
\label{radioObs}

{\it European stations.} Between 2020 Oct 11 and 2021 Feb 04, we
observed \src\ for a total of 68.2~hrs (non-overlapping) with several
European stations that are also part of the European
VLBI\footnote{Very Long Baseline Interferometry} Network (EVN). The
stations involved and the frequencies covered varied between
observations as summarized in Table~\ref{tab:radioObsSummary}. Several
stations observed simultaneously at complementary frequencies in an
effort to cover as large a bandwidth as possible. At each
participating station we recorded the raw voltage data in VLBI Data
Interchange Format \citep[VDIF,][]{2010ivs..conf..192W} with the local
VLBI backends (DBBC2). Each station recorded dual-polarisation
(circular) 2-bit data which were subsequently transferred to Onsala
Space Observatory for processing. We used the same pipeline as
outlined in \citet{kirsten2020} which creates total intensity
filterbanks. Depending on observing frequency, the time and frequency
resolution of the filterbanks are between $64-1024\,\mu$s and
$7.812-2000\,$kHz, respectively.

\begin{table*}
\caption{Summary of the radio observations of \src\ with EVN dishes}
\label{tab:radioObsSummary}
\vspace*{-0.25cm}
\begin{center}
\hspace*{-1.0cm}
\begin{tabular}{lccccc}
\hline
\hline
Station$\mathrm{^{a}}$  & Band$\mathrm{^{b}}$  & Bandwidth [MHz]$\mathrm{^{c}}$ & SEFD [Jy]$\mathrm{^{d}}$ & Fluence limits [Jy$\,$ms]$\mathrm{^{e}}$ & Time observed [hrs] \\
\hline
Wb  & P        & 60       & 2100     & 65      & 34.0\\
O8  & L        & 100      & 350      & 8      &  46.9 \\
Tr  & C     & 250 & 260 & 4  & 3.4 \\
Nt  & L     & 230 & 740 & 12  &  2.0 \\
Mc  & C     & 240  & 170 & 3 &  2.0\\
Ir  & L        & 100      & 700  & 17 &  18.0 \\
Ib  & C, X  & 250, 250 & 620, 650 & 10, 10 & 11.75, 8.60 \\
  \hline
  \multicolumn{5}{l}{Total telescope time/total time on source [hrs]$\mathrm{^{f}}$} & 126.6, 68.2 \\
\hline

\multicolumn{6}{l}{$\mathrm{^{a}}$ Wb: Westerbork RT1, O8: Onsala 25m, Tr: Toru\'n, Nt: Noto 32m, Mc: Medicina 32m, Ir: Irbene 32m, Ib: Irbene 16m} \\
\multicolumn{6}{l}{$\mathrm{^{b}}$ P: 300--364$\,$MHz; L: 1336--1720$\,$MHz; C(Mc,Ib): 4798--5310$\,$MHz; C(Tr): 6500--6756$\,$MHz; X: 8287--8543$\,$MHz} \\
\multicolumn{6}{l}{$\mathrm{^{c}}$ Effective bandwidth accounting for RFI and band edges.} \\
\multicolumn{6}{l}{$\mathrm{^{d}}$ From the \href{http://old.evlbi.org/user_guide/EVNstatus.txt}{EVN status page. SEFD refers to System Equivalent Flux Density.}} \\
\multicolumn{6}{l}{$\mathrm{^{e}}$ Assuming a $7\sigma$ detection threshold} \\
\multicolumn{6}{l}{$\mathrm{^{f}}$ Total time on source accounts for overlap between the participating stations.} \\
\hline
\hline
\end{tabular}
\end{center}
\end{table*}

{\it Deep Space Network.} The Deep Space
Network~(DSN;~\citealt{pearlman+2019a}) is a world-wide array of radio
telescopes that are primarily used for spacecraft communication. The
DSN radio telescopes are located at three main sites (Canberra,
Australia; Goldstone, California; and Madrid, Spain), which each host
a steerable 70\,m radio antenna and several smaller 34\,m radio
dishes. When these radio telescopes are not being utilized to
communicate with spacecraft, they can be used to perform radio
observations of magnetars, high magnetic field pulsars, and fast radio
burst~(FRB) sources (e.g., see~\citealt{Majid+2017, Pearlman+2018,
  Majid+2020, Pearlman+2020b, Pearlman+2020c}).

We performed a continuous radio observation of SGR~1830-0645 using the
Deep Space Network (DSN) 34-m diameter radio telescope (DSS-34),
located at the Canberra Deep Space Communications Complex~(CDSCC) in
Tidbinbilla, Australia, for 5.15 hr starting at 2020 October 12
07:00:01 UTC. The pulsar backend was used to simultaneously record
data at $S$-band (center frequency: 2.2\,GHz, bandwidth: ~118\,MHz)
and $X$-band (center frequency: 8.3\,GHz, bandwidth: ~441\,MHz). Power
spectral density measurements at both frequency bands were channelized
and saved in digital polyphase filterbanks with a frequency and time
resolution of 1\,MHz and 512\,us, respectively. Initial results from
this radio observation were presented in~\citet{pearlman+2020a}.

{\it Green Bank Telescope}. The 100-m Robert C. Byrd Green Bank
Telescope (GBT) observed \src\ for 92 minutes at S-band (center
frequency: 2.0 GHz, bandwidth: 800 MHz) and for 90 minutes at C-band
(center frequency: 6.0 GHz, bandwidth: 1500 MHz) on October 13, 2020,
for a total exposure of 182 minutes using the Versatile GBT
Astronomical Spectrometer (VEGAS) pulsar backend in incoherent
dedispersion and total intensity mode. The S and C bands data were
recorded using 4096 and 3072 frequency channels with a sampling time
of 81.92~$\mu$s and 43.69~$\mu$s, respectively. The data were analyzed
using the
\texttt{PRESTO}\footnote{\url{http://www.cv.nrao.edu/~sransom/presto/}}
software package.

\section{Results}
\label{res}

\subsection{Localization}
\label{local}

We used our \chandra\ observation to derive the \src\ localization. We
selected all events in the 0.5 - 8 keV band detected with the S3
detector of ACIS-S, and used the \texttt{wavdetect} tool of CIAO to
search in the entire field for point sources whose detection
significance is in excess of 5$\sigma$. The search resulted in only
one bright source with a rate of $1.37\pm0.02$~counts~s$^{-1}$
at RA, Dec: 277.673520, -6.754696 degrees, respecitvely, which
corresponds to  RA: 18$^h$30$^m$41.$^s$64, Dec: -06$^\circ$45$'$
16.$\arcsec$9 (J2000). We estimated the positional uncertainty by the
90\% photon containment region around the source position as
0.73$\arcsec$ (statistical only, with another $0.8\arcsec$
  systematic uncertainty). No obvious extended emission is observed
in our short \chandra\ exposure at the arcsecond or arcminute scales.

\subsection{Timing}
\label{timAna}

\begin{figure}[]
  \begin{center}
    \hspace{-0.47in}
    \includegraphics[angle=0,width=0.45\textwidth]{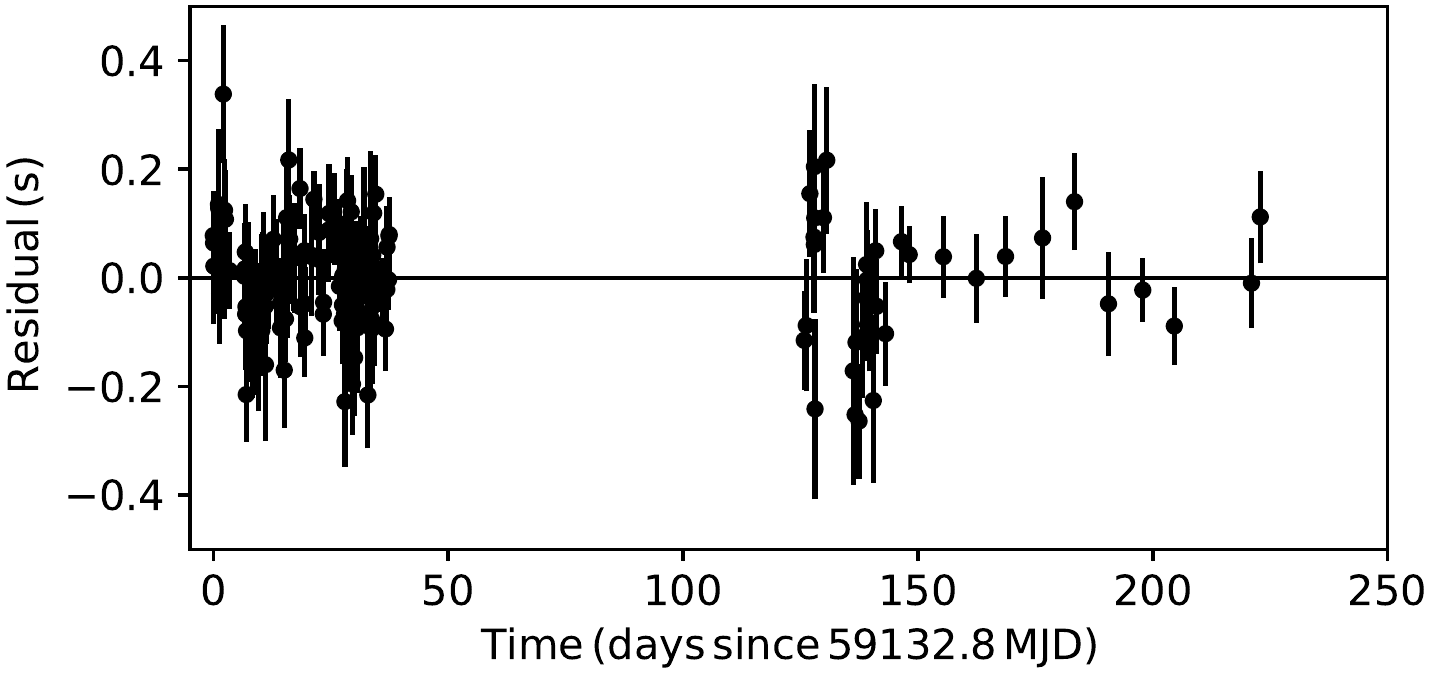}
    \hspace{-0.47in}
\caption{Residuals, in seconds, of a timing model fit to the \src\
  TOAs as derived from 400 to 1200 seconds of \nicer\ data. The gap constitutes the sun-constrained period. The model includes up to the fourth term in the Taylor expansion of the phase evolution, and it is summarized in Table~\ref{timDat}.}
\label{fig:resids}
\end{center}
\end{figure}

To develop a phase-connected timing model we computed a set of pulse
times of arrival (TOAs) from all \nicer\ observations and fit them to
a model. We compute TOAs using an unbinned maximum likelihood
technique described in \citet{ray11ApJS}. This technique uses an
analytic pulse profile template in the likelihood calculation. Because
the main pulse changes from double peaked early in the outburst
(around MJD 59154) to a single peak later, we used two templates. We
tried several different template alignments, i.e., a single wide
Gaussian, a multi-gaussian model, and aligning the fundamental of a
Fourier series fit to the pulse profile. We chose the ones where
the timing residuals showed the smallest discontinuity at the
switchover date. This turned out to be a single wide Gaussian, with
slightly different widths for the early and late data (FWHM = 0.39 and
0.32 respectively). Using these templates, we computed TOAs for each
segment of data of typical duration 400--1200 s. Using PINT
\citep{luo19ascl:pint}, we fit the TOAs to a timing model of the form
$\phi(t)=\phi_0+\nu(t-t_0)+1/2\dot{\nu}(t-t_0)^2+1/6\ddot{\nu}(t-t_0)^3+\ldots$,
truncated at the highest significant (at the $5\sigma$ level)
term, which in this case was $\dddot{\nu}$. Given the high \nicer\
cadence, we were able to maintain phase coherence throughout our
monitoring campaign. The parameters of the best-fit timing model are
presented in Table~\ref{timDat} and the residuals are displayed in
Figure~\ref{fig:resids}. We find that the \src\ spin frequency is
$\nu=0.096008680(2)$~Hz and spin-down rate is
  $\dot{\nu}=-6.2(1)\times10^{-14}$~Hz~s$^{-1}$, which imply a dipole
field strength at the equator $B\approx2.7\times10^{14}$~G, a spin-down age
$\tau\approx24.4$~kyr, and a spin-down luminosity
$|\dot{E}|\approx2.4\times10^{32}$~erg~$s^{-1}$.

\begin{deluxetable}{lr}
\tablecaption{\src\ Timing Parameters
\label{timDat}}

\tablehead{\colhead{Parameter} & \colhead{Value}}
\startdata
R.A. (J2000) & 18:30:41.64 \\ 
Decl. (J2000) & $-6$:45:16.9 \\
Time Scale & TDB \\
Ephemeris & DE405\\
Epoch (MJD) & 59132.0 \\
$\nu$ (Hz)  & 0.096008680(2) \\
$\dot{\nu}$ (Hz s$^{-1}$) &  $-6.2(1)\times10^{-14}$ \\
$\ddot{\nu}$ (Hz s$^{-2}$) & $4.7(3)\times 10^{-21}$ \\
$\dddot{\nu}$ (Hz s$^{-3}$) & $-2.7(4)\times 10^{-28}$ \\
Valid Range (MJD) \mbox{\hspace{32pt}} &  59132.7--59355.7 \\
$\chi^2$/dof  &  177/175 \\
RMS residual (ms)  &  95 \\
\hline
\multicolumn{2}{c}{Inferred Parameters} \\
\hline
$B$ (G)  & $2.7\times10^{14}$\\
$\tau$ (kyr) & 24.4 \\
$|\dot{E}|$ (erg s$^{-1}$)  & $2.4\times10^{32}$\\
\enddata
\end{deluxetable}

\begin{figure*}[th!]
\begin{center}
  \includegraphics[angle=0,width=0.311\textwidth]{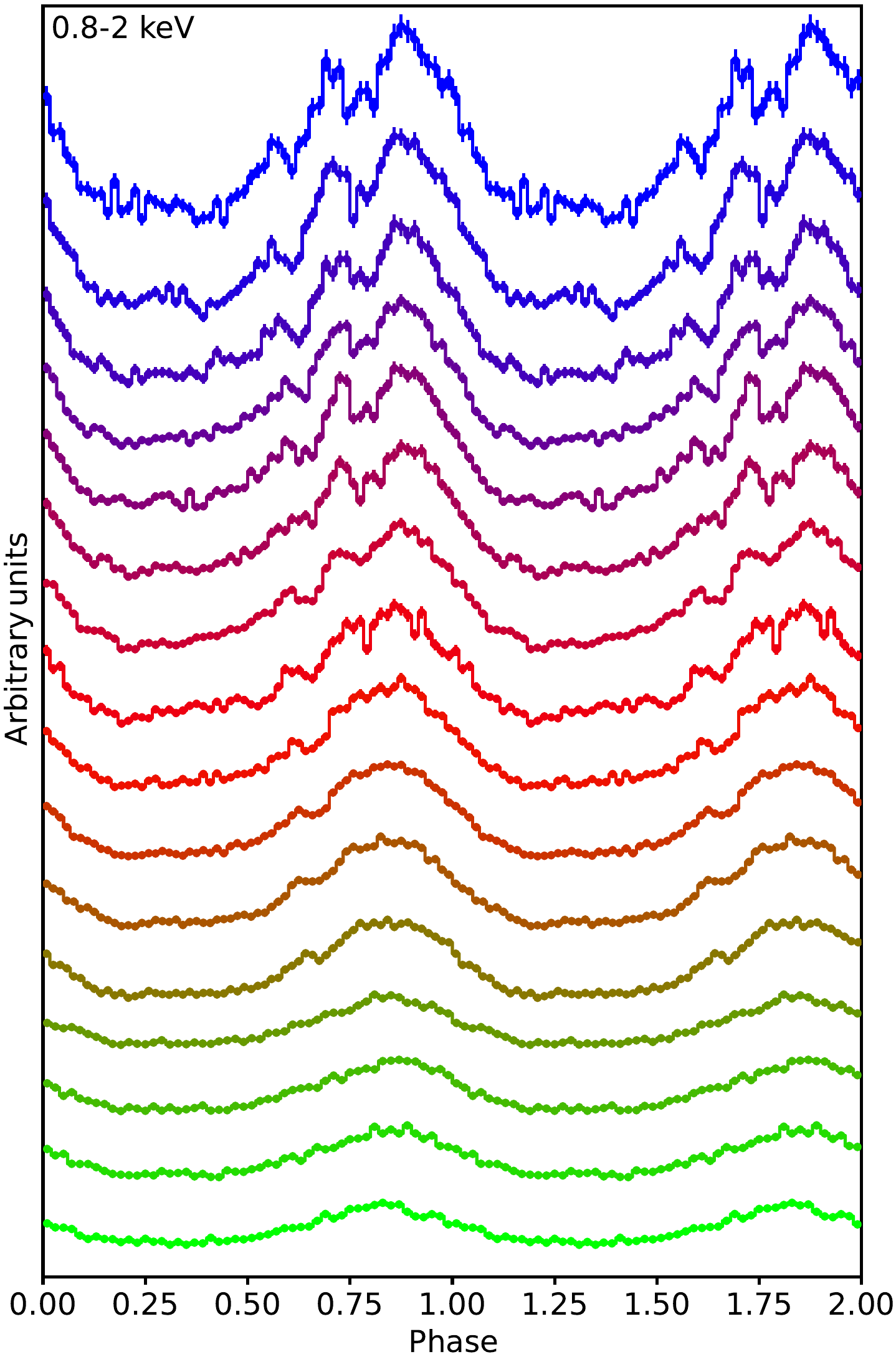}
  \includegraphics[angle=0,width=0.311\textwidth]{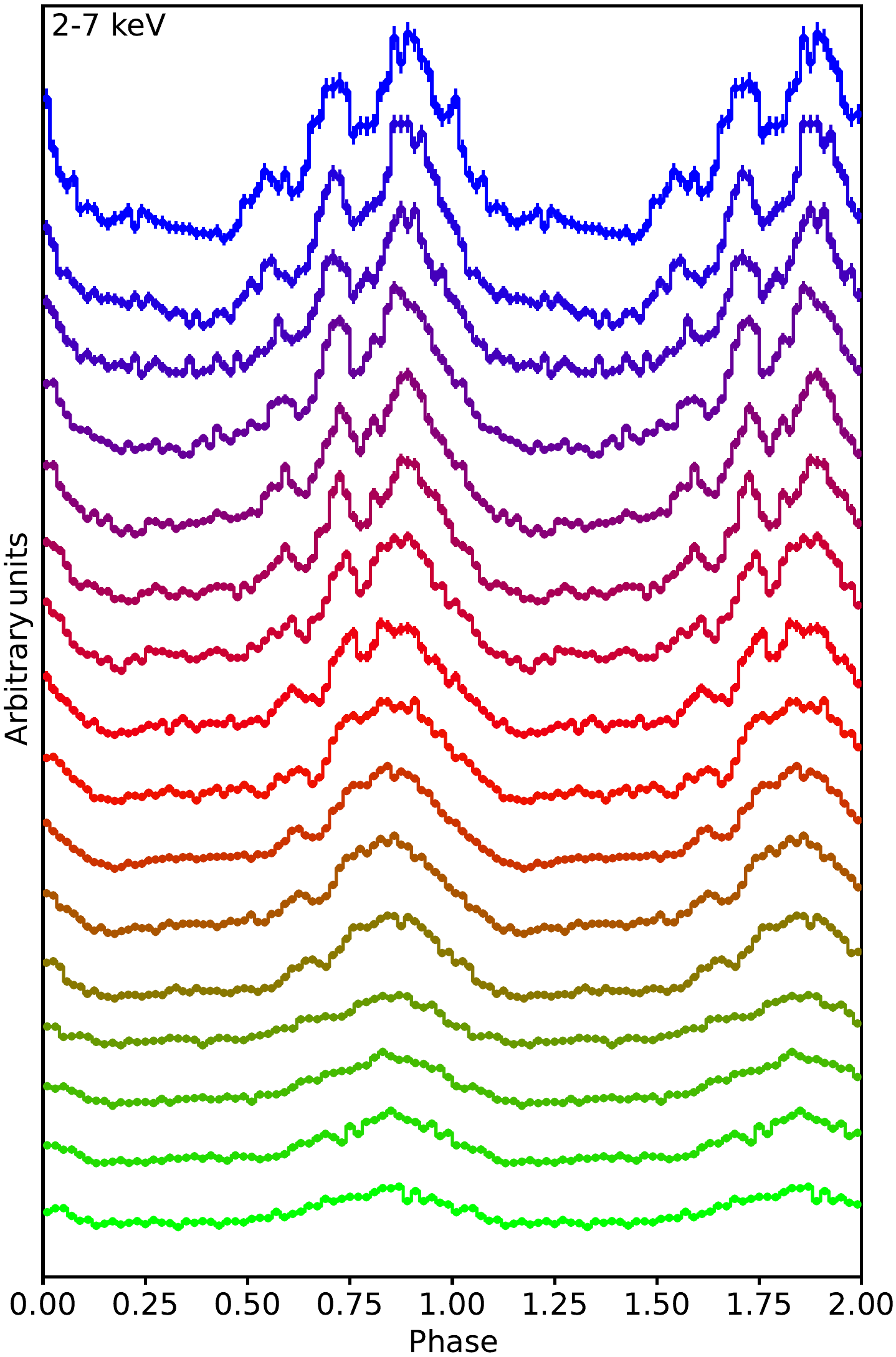}
  \includegraphics[angle=0,width=0.362\textwidth]{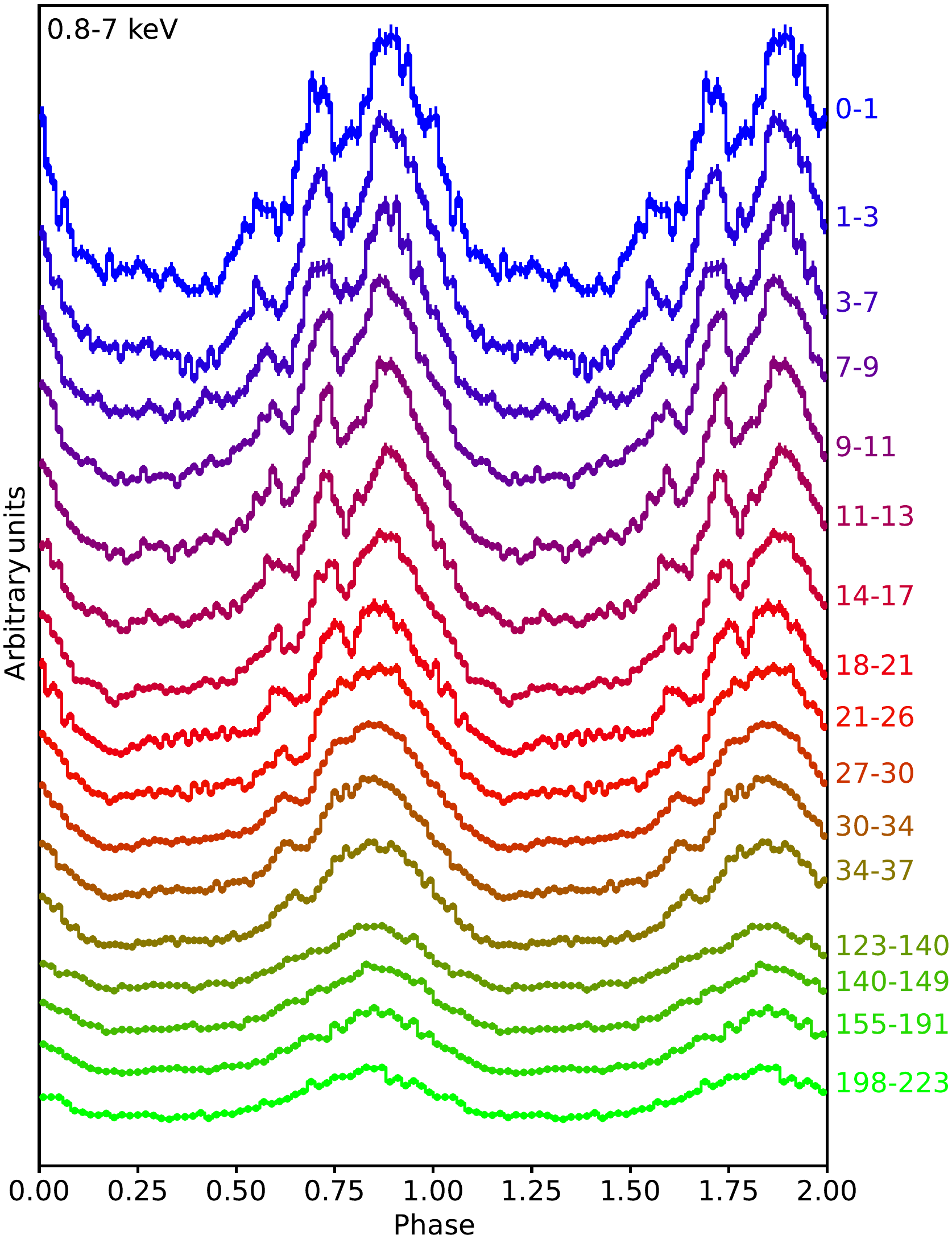}
\caption{\src\ persistent emission pulse profile evolution with time
  in the 0.8--2~keV ({\sl left-panel}), 2--7~keV ({\sl middle-panel}),
  and 0.8--7 keV range ({\sl right-panel}). All profiles are corrected
  for the background. Colors represent the days during which the data
is accumulated post 2020 October 10. Two rotational cycles are
shown. The y-axis is shifted arbitrarily for clarity. See text for
more details.}
\label{ppTimRes}
\end{center}
\end{figure*}

The time-integrated soft X-ray pulse profile of \src\ initially shows
a complex shape, with 3 prominent peaks, each separated by
$\approx0.2$~cycles. We searched for any temporal evolution by
building pulse profiles using observations spanning one to several
days, accumulating for each profile an exposure ranging from 4~ks to
40~ks. The right panel of Figure~\ref{ppTimRes} displays the
0.8--7~keV pulse profiles with 70 phase bins color-coded by time
interval in days from MJD 59132.0. The prominent peaks are clearly
visible up to day 21 post outburst, after which the profile simplifies
to a prominent single peak. However, the fainter peak around
rotational phase of 0.6 remained visible up until the last observation
prior to the sun-constraint period, 37 days post-outburst. We note a
clear decrease in the separation between the peaks as the outburst
evolves. A more detailed analysis of this behavior will be presented
elsewhere. The pulse profiles following the sun-constrained period
clearly exhibit a simple broad single-peak pulse.

We also constructed pulse profiles in two energy bands, namely
0.8--2~keV and 2--7 keV (Figure~\ref{ppTimRes}, left and middle
panels, respectively). These energy ranges reflect the contribution
from the two thermal models necessary to describe the 0.8--7 keV
spectrum (see Section~\ref{specAna}). In both energy ranges, the three
peaks are present, and at similar phases. The only subtle distinction
in the profiles is the dip between the peaks, where it is more
prominent at hard X-rays.

\begin{figure}[]
  \begin{center}
    \includegraphics[angle=0,width=0.49\textwidth]{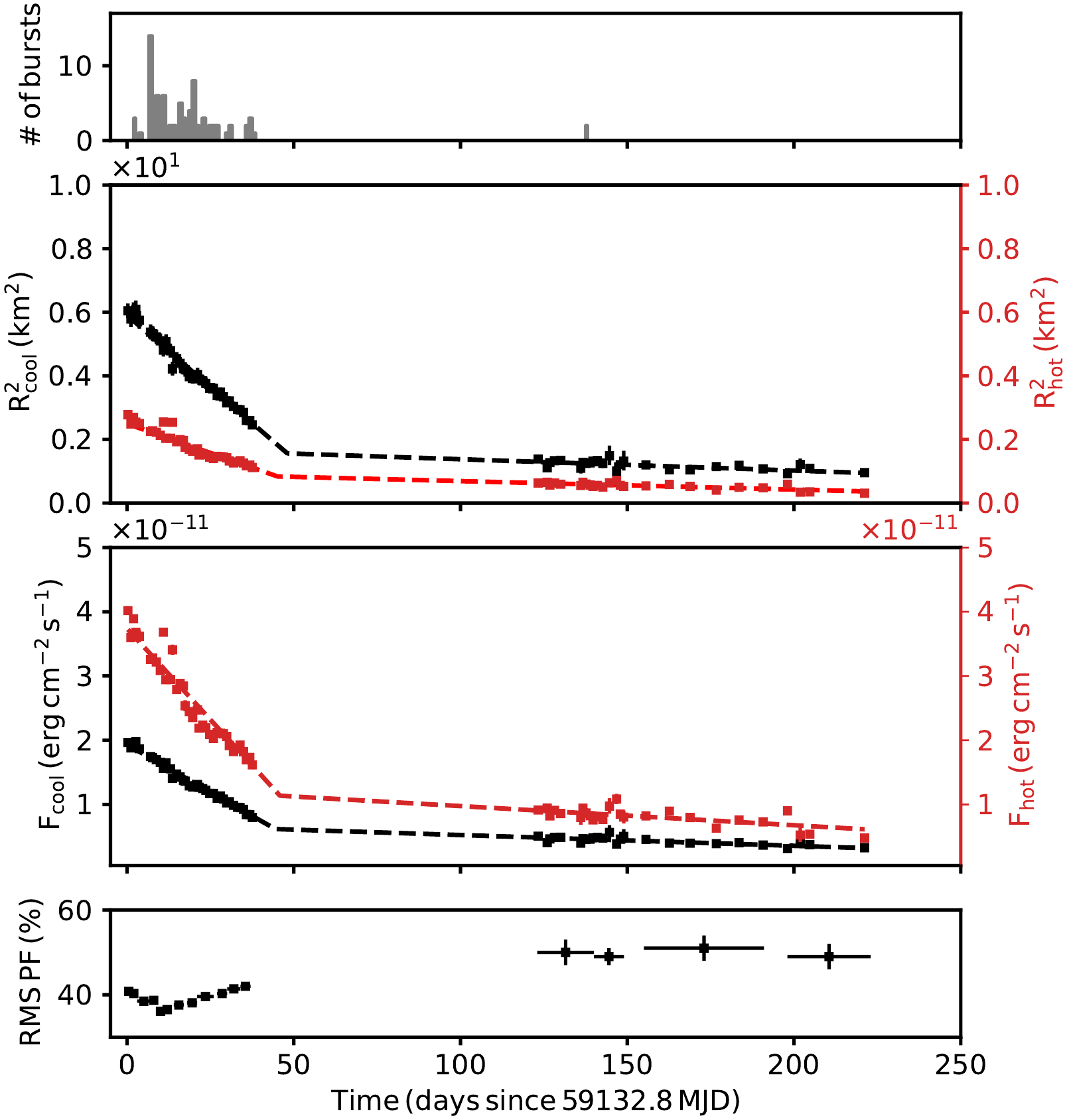}
\caption{{\sl Panel (a).} Number of \src\ X-ray bursts per day as
  observed by \nicer. {\sl Panel (b).} Decay of the hot (red squares)
  and warm (black squares) BB emitting areas throughout the source
  outburst. {\sl Panel (c).} Decay of the absorption-corrected
  flux of the hot (red squares) and warm (black squares) BB components
  in the 0.5-10~keV range. In panels (b) and (c) the dotted lines are
  the best fit broken-linear decay function. {\sl Panel d}. RMS pulsed
  fraction evolution with time. See text for more details.}
\label{fig:spEv}
\end{center}
\end{figure}

We fit the background-corrected pulse profiles to a Fourier
series to measure the root-mean-square pulsed fraction (${\rm
  PF_{rms}}$, \citealt{woods04ApJ:1E2259}). We consider the smallest
number of harmonics that provide the best fit to each profile
according to an F-test. We find that the first seven harmonics are
required to provide an adequate fit to the early pulse shapes. Yet,
three harmonics are adequate to result in
satisfactory fit to the data post sun-constrained period. The source
exhibits a highly pulsed signal, with ${\rm PF_{rms}}$ narrowly
varying between $36\%$ and $42\%$ during the first month of the
outburst (bottom panel of Figure~\ref{fig:spEv}). Subsequent to the
sun-constrained period, the source pulse fraction increased to
$\approx50$\%. We find no significant variation in the rms PF with
energy.

\begin{figure}[t]
\begin{center}
\hspace{-0.2in}
  \includegraphics[angle=0,width=0.49\textwidth]{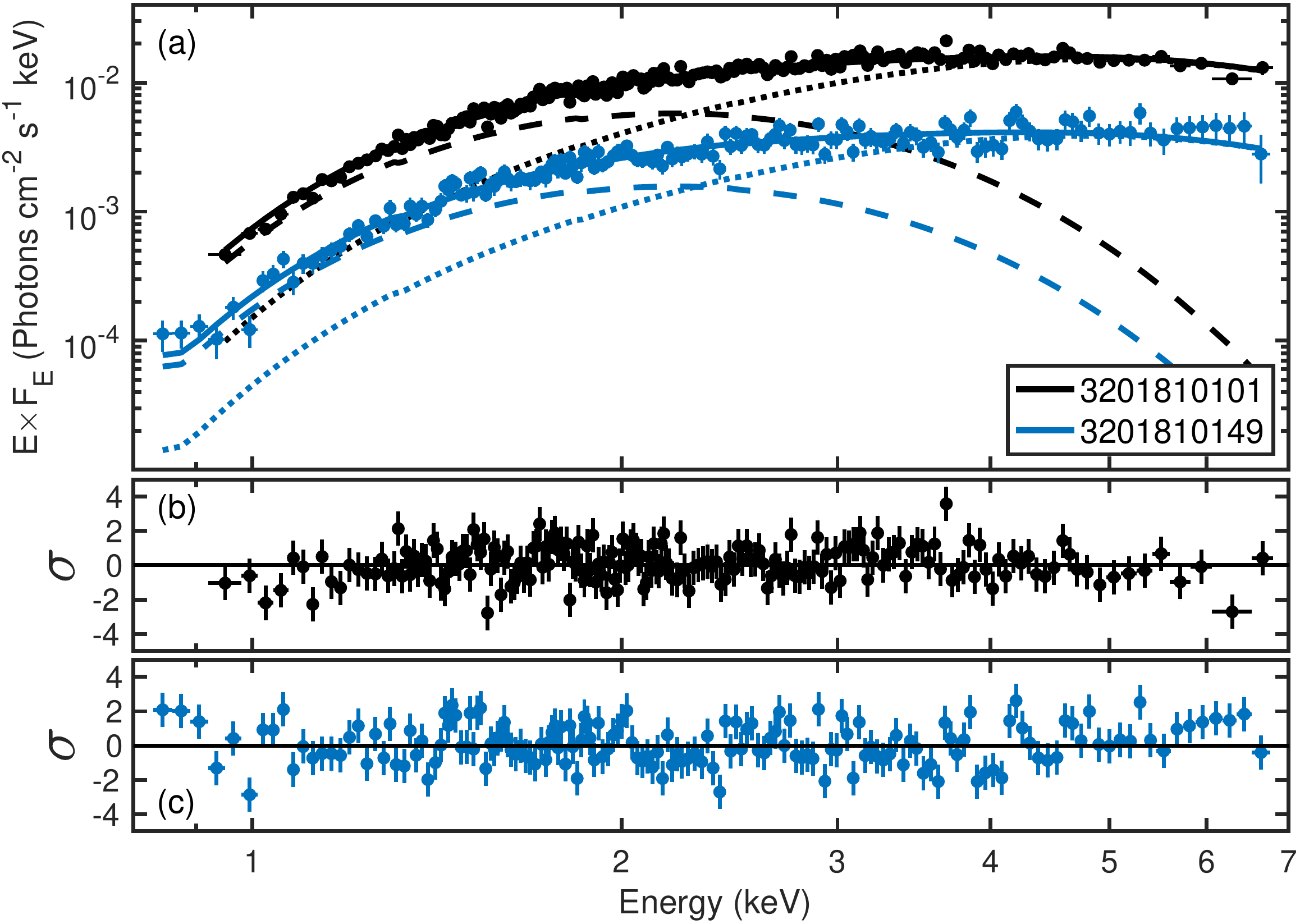}\\
\caption{{\sl Panel (a).} Example \nicer\ spectra from the first observation post outburst (4 hours after the source discovery, black dots) and 138 days later (blue dots). The dashed and dotted lines are the best fit absorbed warm and hot BB models, respectively, while the solid line is their sum. {\sl Panels (b) and (c)}. Residuals, in units of standard deviation $\sigma$, of each spectrum from the best fit absorbed 2BB model.}
\label{specExamp}
\end{center}
\end{figure}

\subsection{Persistent Emission Spectroscopy}
\label{specAna}

We fit the 0.8--7~keV X-ray spectra extracted from all 37 observations
taken prior to the sun-constrained period simultaneously, with an
absorbed two blackbody (BB) model. We link the hydrogen column density
among all the spectra since we do not expect any large and variable
absorption intrinsic to the magnetar. Initially, we let all other
parameters free to vary, i.e., the temperatures and size of the
apparent emitting area. We find a reduced $\chi^2$ between 0.9 and 1.1
for all spectra, which have degrees of freedom (dof) ranging from
  40 to 440. We observe strong variations in the size of the apparent
emitting areas for the warm and hot BB components. The temperatures on
the other hand remain essentially constant through all observations,
varying only at $<2\sigma$ level. Hence, we linked the temperatures of
both components between all the spectra and refit the data. This
resulted in an equally good fit with for all spectra. We find a
hydrogen column density in the direction of the source $N_{\rm
  H}=(1.17\pm0.01) \times 10^{22}$~cm$^{-2}$, while the temperatures
of the warm and the hot BB are $0.47\pm0.01$~keV and
$1.22\pm0.01$~keV, respectively. We also fit these spectra with
  the sum of a BB and a power-law, affected by interstellar
  absorption. This model consistently provide a worse fit compared to
  the 2BB with $\chi^2$ larger than that of the latter for the same
  number of dof. This is consistent with the 0.5-30~keV broad-band
  spectral result presented in \citet{cotizelati21ApJ1830} which found
  that a purely thermal model is the best fit to the soft X-ray part
  of the spectrum. Hence, for the remainder of the spectral analysis,
  we adopt the 2BB model.

We then fit the post sun-constrained period spectra simultaneously
with an absorbed thermal model. We fixed the hydrogen column density
to the value as derived above. We find that a single thermal component
cannot adequately describe the spectral shape of these spectra, and a
2BB model is still required. We started by allowing the areas and
temperatures to vary freely. Once again, the temperatures of all
spectra were consistent within uncertainties, hence, were linked. This
final fit resulted in a reduced $\chi^2$ of about 1 for all
spectra. The temperature of the warm and hot BB components are
$0.48\pm0.02$~keV and $1.22\pm0.05$~keV,
respectively. Figure~\ref{specExamp} shows the $E F_{\rm E}$ spectra
extracted from a pre- and post-sun-constrained observation in black
and blue, respectively, demonstrating that the flux decay lacks
significant spectral shape evolution.

The temporal evolution of the emitting area, $R^2$ and the
absorption-corrected 0.5-10~keV
flux of each BB component is shown in Figure~\ref{fig:spEv}. Each of
these parameters for both BB components follow a similar trend, and
are well fit with a broken first-order polynomial function
(dashed-lines). The fits result in $\chi^2$ of 89, 96, 77, 105,
  for 59 dof, for the warm and hot areas and corresponding fluxes,
  respectively. The break occurs shortly after the start of the 
sun-constrained period at $t_{\rm break}=46\pm2$~days or around
2020 November 27. We find a flux decay rate of
$-5.7(2)\times10^{-13}$~erg~s$^{-1}$~cm$^{-2}$~day$^{-1}$ and 
$-3.0(5)\times10^{-14}$~erg~s$^{-1}$~cm$^{-2}$~day$^{-1}$, before and
after $t_{\rm break}$, respectively. The ratio of the BB
fluxes shows no apparent variation with an average $<F_{\rm
  hot}/F_{\rm warm}>\approx1.9\pm0.2$. The areas of the hot and warm
components shrank from 0.28~km$^2$ to 0.04~km$^2$ and 6.0~km$^2$ to
1.0~km$^2$, respectively.

\subsection{X-ray Bursts}

\begin{figure}[]
  \begin{center}
    \hspace{-0.47in}
    \includegraphics[width=0.47\textwidth]{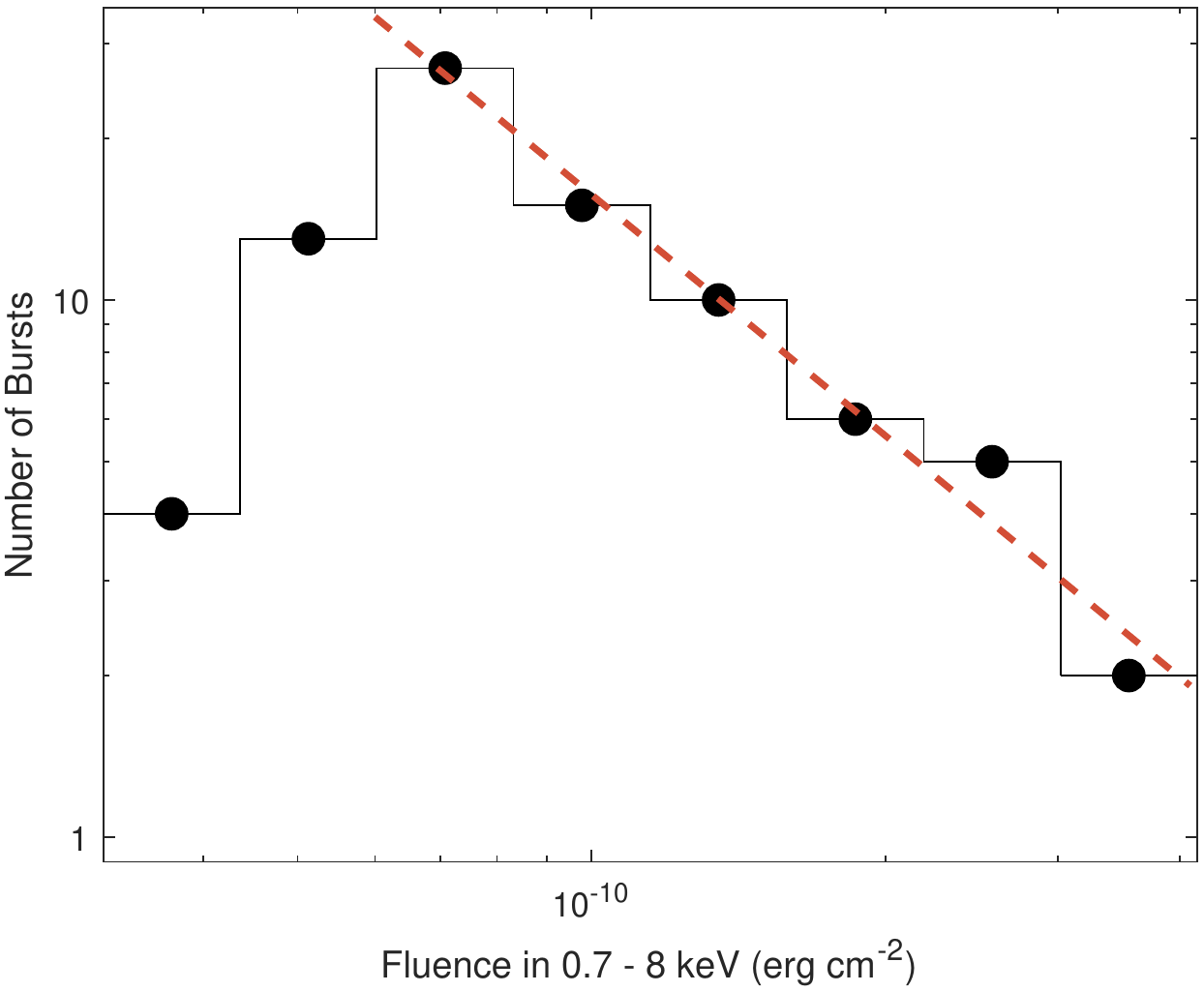}
    \hspace{-0.47in}
\caption{Fluence distribution of bursts in SGR 1830$-$0645 detected
  with \nicer. For fluences $\gtrsim6\times10^{-11}$~erg~cm$^{-2}$,
  the distribution approximately follows $F^{-0.5}$, as shown with
  a red, dashed line.}
\label{fig:fluence_distribution}
\end{center}
\end{figure}

We searched for bursts in the X-ray band with the \nicer\ data sets in
the energy range of $0.7$-$8$ keV. We use the Bayesian block technique
to search for significant X-ray flux variability in individual event
files \citep{scargle2013apj:BB}. The blocks with duration shorter than
$\sim$1 s are further examined to exclude those with high backgrounds
and multiple blocks in one burst. Moreover, blocks close to the
boundaries of a GTI have also been excluded. The detection
significance of a burst is defined through the Poisson probability of
detecting a number of photons in a block, given an estimated non-burst
count rate calculated from nearby 1~s intervals. We have identified 84
short bursts with Poisson probabilities lower than
$2.6\times10^{-7}$. This corresponds to a $5.2~\sigma$ detection
significance, and implies a false alarm rate of $\le1$ throughout our
campaign.

Among the 84 bursts, the averaged Bayesian block duration is
about 0.029~s with a standard deviation of 0.023 s. The average photon
count rate for the burst ensemble is $470\pm14$ counts s$^{-1}$,
corresponding to an unabsorbed flux of
  $(4.5\pm0.2)\times10^{-9}$~\fluxcgs in the energy range of 0.7--8
keV (assuming a blackbody with a temperature of 1.5 keV; as derived
through our burst spectral analysis below). The burst with the highest
fluence occurred at MJD (TDB) 59139.864807 with a duration of 0.014 s
and an averaged flux of $3.3\pm0.5\times10^{-8}$~\fluxcgs. The
fluence distribution of the bursts is shown in Figure
\ref{fig:fluence_distribution}. The high-fluence tail, which covers
the range 6--30 $\times10^{-11}$~erg~cm$^{-2}$, can be described by a
power-law with an index of $\approx1.5$. To estimate the uncertainty
of this index, we performed $10^5$ times of Monte Carlo simulations.
In each simulation, we generated Poisson distributed photons in each
burst with a mean value equal to the observed one, and calculated the
corresponding fluence. The result suggests an uncertainty on the PL
index of 0.3. 

\begin{figure}[]
  \begin{center}
    \hspace{-0.47in}
    \includegraphics[angle=0,width=0.49\textwidth]{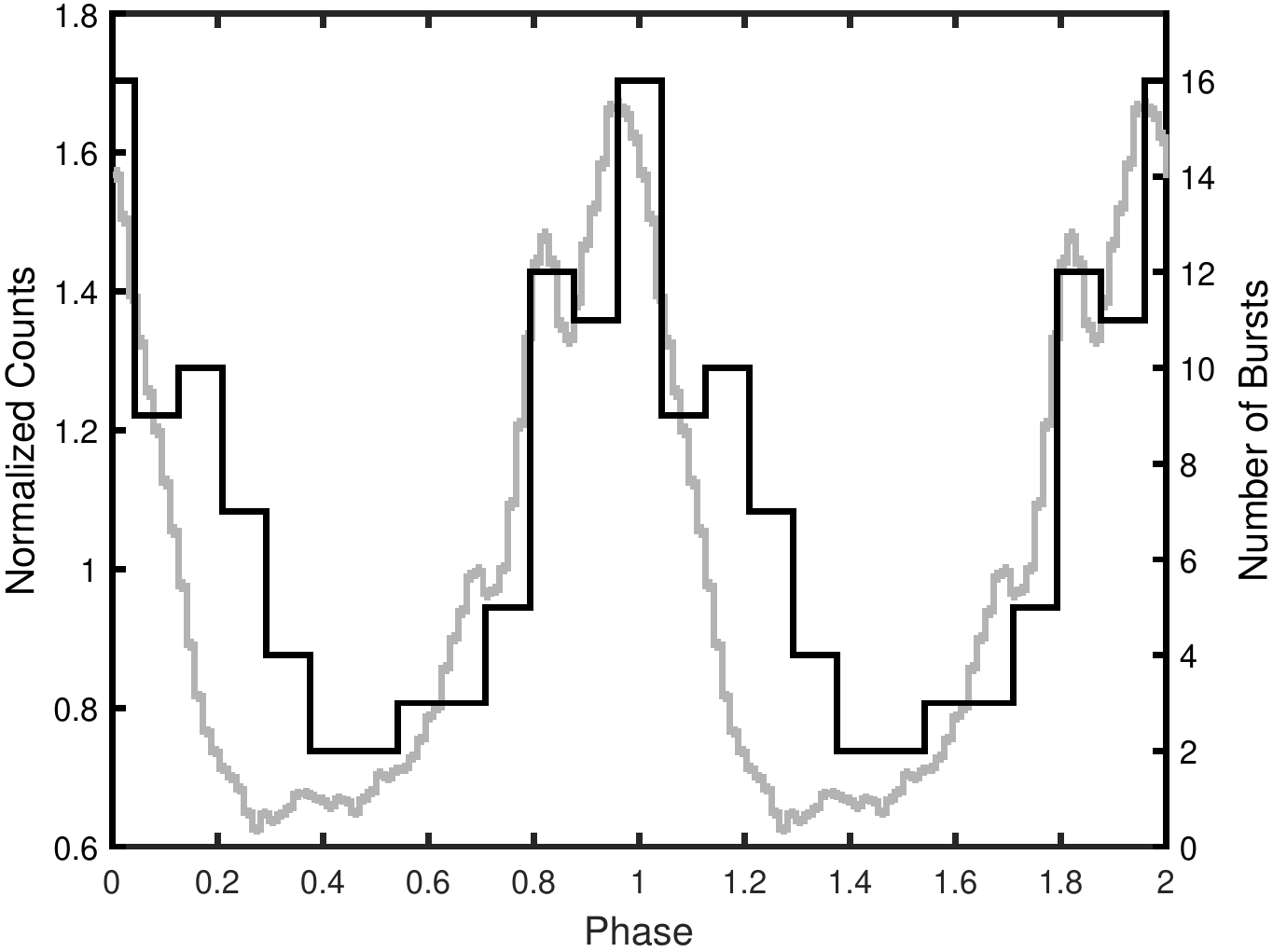}
    \hspace{-0.47in}
\caption{Average pulse profile of \src\ (gray) in the energy range 0.8--7 keV, using all X-ray observations reported in this paper. The black profile represents the histogram of the bursts phase folded using the spin parameters as reported in Table~\ref{timDat}.}
\label{fig:bursts}
\end{center}
\end{figure}

We investigate the spin phase distribution of these short bursts using
the timing model presented in
Table~\ref{timDat}. Figure~\ref{fig:bursts} shows the burst phase
distribution along with the pulse profile of \src\ as derived from all
observations. It is clear that the burst phase distribution is not
uniform, with its peak occurrence coinciding with the peak of the
persistent emission X-ray pulse profile. We apply the Anderson Darling
test to assess the burst phase against a uniform distribution which
yielded an AD statistic of 4.5 with a corresponding p-value of
0.005. Moreover, we performed simulation to test the significance of
this apparent non-uniformity in phase. In each simulation, we generate
83 mock bursts located at randomly distributed spin phases. For $10^5$
trials, only $\approx450$ of them result in AD statistics higher than
the observed value. This implies that the null hypothesis of the
observed burst phase distribution being drawn from a uniform
distribution can be rejected at the $3\sigma$ level. Finally, we
searched for any phase dependence of the mean photon energy and the
fluence of bursts, but found no clear trend.

\begin{figure}[]
  \begin{center}
    \hspace{-0.47in}
    \includegraphics[width=0.49\textwidth]{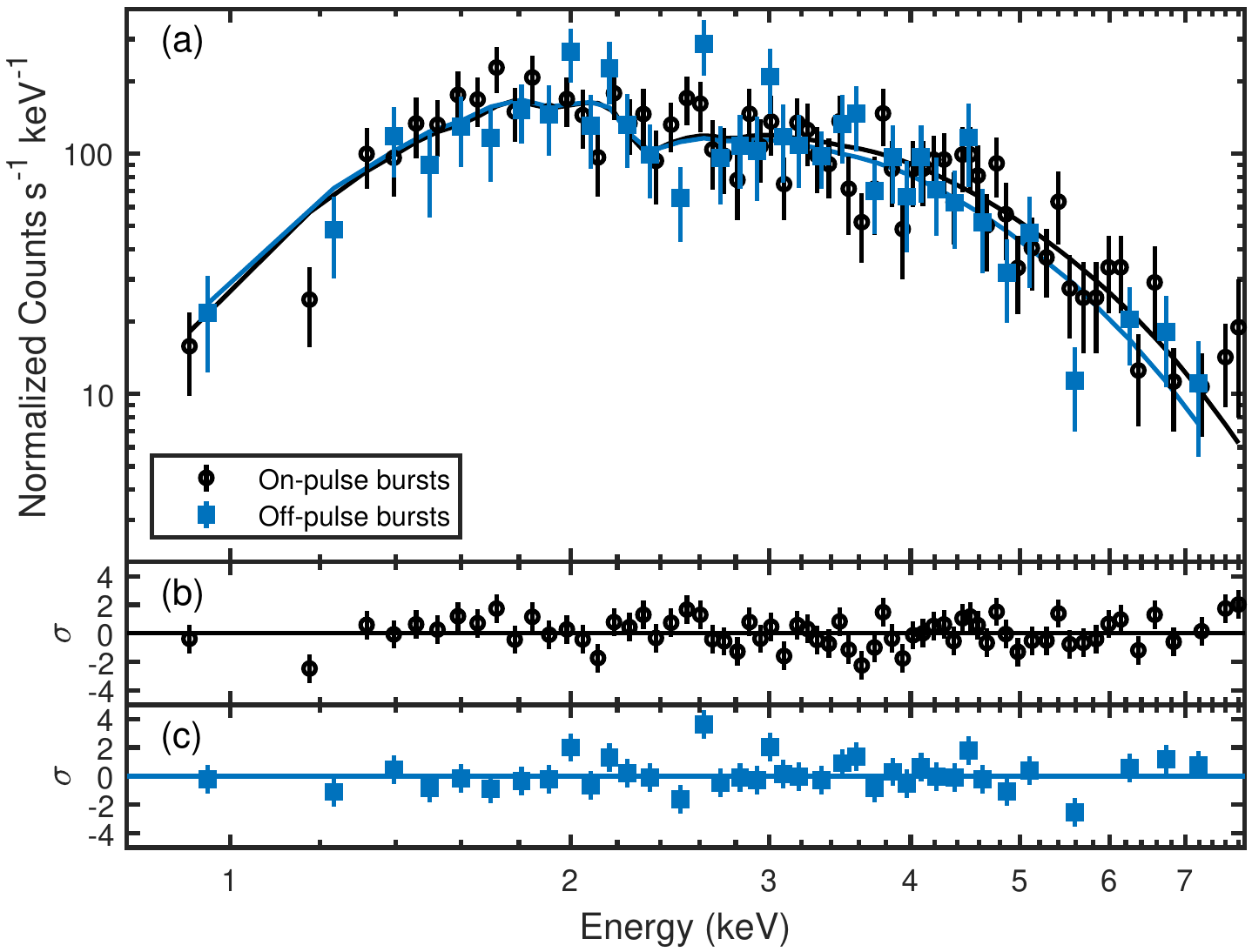}
    \hspace{-0.47in}
\caption{(a) The \nicer\ spectra of the accumulated on-pulse bursts (black circles) and off-pulse bursts (blue squares). The black and blue lines represent the best-fit power-law model. (b) and (c): Residuals of each spectrum from the best-fit power-law model in terms of $\sigma$.}
\label{fig:burst_spectrum}
\end{center}
\end{figure}

We accumulated a total of 1111  bursts photons in the 0.7--8
keV~range. We merged photon events with \texttt{nimpumerge} and
generated an averaged burst spectrum in the burst time intervals. We
fit the spectrum with the Cash statistics \citep{Cash1979}, assuming
that the background constitutes the non-burst epochs. We excluded the
data-sets taken in 2021 due to the dramatic decrease in the rate of
bursting activity. The burst spectrum can be described equally well
with an absorbed PL or BB model. We fix the hydrogen column
  density $N_{\rm H}$ to $1.17\times10^{22}$~cm$^{-2}$ as derived
  through the persistent emission spectral analysis. We find a photon
index of $0.4_{-0.3}^{+0.5}$ (C$_{\rm stat}=408.5$ with 448 degrees of
freedom) for the PL component, or $kT=1.5\pm0.1$ keV and a radius of
$4.1\pm0.3$ km ($C^2=362.6$ with 448 degrees of freedom) when
considering the BB model. We further divided bursts into two groups:
bursts associated with the X-ray pulse (on-pulse, phase 0.8--1.3) and
the X-ray valley (off-pulse, phase 0.3--0.8). No significant
difference is obtained between the spectra of these two groups. The
on-pulse and off-pulse spectra are shown in Figure
\ref{fig:burst_spectrum}.

\subsection{Radio upper-limits}

{\it Deep Space Network.} We dedispersed the $S$-band and $X$-band
data from DSS-34 with trial DMs between 10 and
2000~pc~cm$^{\text{--3}}$ and independently searched the data in
each frequency band for periodic radio emission. We used a
GPU-accelerated Fast Folding Algorithm~(FFA) to search for pulsed
radio emission with periods between 9 and 12\,s. We also used a
GPU-accelerated Fourier Domain Acceleration Search (FDAS) pipeline,
which employs a matched filtering algorithm to correct for Doppler
smearing, to search for periodicity. No statistically significant
periods, with a signal-to-noise~(S/N) ratio above 7.0, were found
after folding the data modulo each of the period candidates identified
by the two algorithms. We also folded the data utilizing the
  \nicer\ ephemerides at the DSN epoch, as well as using trial periods
  around the predicted 10.42~s rotational period of the source. We do
  not find any pulsed emission at or near the expected signal. For
an assumed duty cycle of 10\%, we place the following (7-sigma) upper
limits on the SGR~1830-0645 radio flux density during our radio
observation: $<$\,0.14\,mJy at $S$-band and $<$\,0.08\,mJy at
$X$-band.

We also searched the de-dispersed $S$-band and $X$-band data for radio
bursts using a Fourier domain matched filtering algorithm, where each
dedispersed time-series was convolved with boxcar functions with
logarithmically spaced widths between 512\,us and
153.6\,ms. Candidates with detection signal-to-noise ratios above 7.0
were saved and classified using a GPU-accelerated machine learning
pipeline based on the FETCH (Fast Extragalactic Transient Candidate
Hunter) software package~\citep{Agarwal2020MNRAS}. The dynamic spectra
of the candidates were also visually inspected for verification. No
radio bursts were detected above a 7-sigma fluence detection threshold
of 1.8\,Jy\,ms for a 1-ms burst at $S$-band and 1.0\,Jy\,ms for a 1-ms
burst at $X$-band.

{\sl European stations.} We searched the filterbanks from each station
for single pulses using
Heimdall\footnote{\href{https://sourceforge.net/projects/heimdall-astro/}{https://sourceforge.net/projects/heimdall-astro/}}
in the DM-range $\mathrm{10-2000\,pc\,cm^{-3}}$. Classification of
candidates was performed via FETCH \citet{Agarwal2020MNRAS} and manual
inspection of the most promising candidates. No attempt of folding the
data around the rotational period was done. We found no bursts above a
S/N threshold of 7 in any of the observations, constraining the
fluence of any potential bursts as summarized in Table
\ref{tab:radioObsSummary}.

{\it Green Bank Telescope}. The data were searched for periodic
emission, both completely blind and by folding the data around the
known period, and for bright single pulses in the DM range of
10--1500~pc~cm$^{-3}$. We do not detect any significant periodic or
single-pulse emission. For periodic emission, we place a 7-sigma upper
limit on the average flux density of 7.8~$\mu$Jy and 6.2~$\mu$Jy at S
and C band respectively, assuming a 10\% duty cycle. The dedispersed
data were also searched for fast radio bursts using Heimdall and
candidates were run through the GPU accelerated convolutional neural
network FETCH \citep{Agarwal2020MNRAS} to distinguish between radio
frequency interference (RFI) and astrophysical signals. We did not
detect any radio bursts in our data and the 7-sigma fluence upper
limits are 0.07 Jy~ms and 0.04 Jy~ms at 2.0 GHz and 6.0 GHz,
respectively, assuming a pulse width of 1~ms.

\section{Summary and discussion}
\label{discuss}

In this paper, we present the results of a \nicer\ heavy-cadence
monitoring campaign of the recently discovered magnetar \src, covering
the time interval from 4 hours to 223 days post outburst-onset. This
dense observing campaign allowed us to track the X-ray spectral and
temporal evolution of the source in exquisite detail, while affording
an analysis of low-level bursting activity that occurred throughout
the outburst. We complement \nicer\ with a \chandra\ observation that
provided the precise sky localization of the source. Finally, we
present the results of a suite of \src\ radio observations taken at
several frequencies from multiple epochs during the source outburst.

\subsection{General properties}

The \nicer\ observations of \src\ allowed us to build an accurate
timing solution to the rotational properties of the source over
several months. The temporal characteristics we inferred, $\nu=0.096008680(2)$~Hz and
  $\dot{\nu}=-6.2(1)\times10^{-14}$~Hz~s$^{-1}$, are quite typical of
the magnetar family, as well as the derived intrinsic properties:
$B=2.7\times10^{14}$~G, $\tau=24.4$~kyr, and
$|\dot{E}|=2.4\times10^{32}$~erg~$s^{-1}$ \citep[see
also][]{cotizelati21ApJ1830}. The long baseline of the
observations also revealed the presence of higher order frequency
derivatives, likely indicating a significant level of timing noise
\citep[typical of magnetars during outburst epochs,
e.g.,][]{younes15ApJ:1806,archibald20ApJ:1048,hu20ApJ:1818}. All the
above properties, along with the strong bursting activity and the
persistent flux decay, cement the origin of \src\ as the latest
addition to the magnetar family.

The Galactic coordinates of \src\ (section~\ref{local}) locate this
magnetar close to the positions of known magnetars 
SGR~1833$-$0832, Swift J1834.9$-$0846, and 1E~1841$-$045. These three
sources are thought to be on the Galactic Scutum–Centaurus Arm (Table
7 and Figure 17 of \citealt{2017ApJS..231....8E}). The distances of
the latter two sources, Swift J1834.9$-$0846, and 1E~1841$-$045, are
measured at 4.2~kpc and 8.5~kpc from the associated supernovae
remnants (SNRs) W41 \citep{leahy08AJ:w41} and Kes~73
\citep{tian08ApJkes73}, respectively. Due to the lack of any SNR
around \src, we assume a fiducial distance of \src\ at 4~kpc as the
close part of this spiral arm. The \chandra\ observation reveals no
significant extended emission around the source that may be
interpreted as an X-ray wind nebula, in-line with the majority of the
magnetars (with the exception of Swift J1834.9$-$0846,
\citealt{younes16ApJ:1834}).

\subsection{X-ray evolution during outburst}

\src\ pulse profile at the onset of the outburst shows a complex
morphology, exhibiting three clearly distinct peaks. The initial
separation of the peaks is about 0.2 rotational phases, which seem to
decrease with time. By day 21 post-outburst, the profile simplified to
a broad single peak, though the weakest peak around phase 0.65
remained visible until the last observation prior to the
sun-constraint period. Complex profile shapes are quite common for
magnetars during their outburst epochs. For instance, 4U 0142+61
double-peaked profile revealed extra peaks following the source 2006
and 2015 outbursts \citep[][see also, e.g.,
\citealt{gavriil11ApJ:0142,
  rea13ApJ:0418}]{archibald17ApJ:0142}. These changes commonly revert
back to a simpler shape as the outburst wanes. This is also the case
for \src, where the pulse profiles following the sun-constrained
period, i.e., four months post-outburst, evince a single broad pulse.

We find no strong energy dependence on pulse shape, except for a
deeper, more pronounced trough between the two main peaks in the
energy range 2--7~keV compared to the lower 0.8--2~keV band. Yet,
these two energy ranges are dominated by each of the two BB components
used to fit the 0.8--7 keV phase-averaged spectrum
(Figure~\ref{specExamp}). These two BB components evolve similarly
throughout the outburst, i.e, their temperatures remain constant while
their areas decrease, driving a very similar flux decay trend
(Figure~\ref{fig:spEv}). Hence, it is evident that these two spectral
components are not distinct. It is plausible that the X-ray emitting
area is not isothermal, but rather possesses a temperature gradient,
for example a hot core surrounded by a cooler ring, which are evolving
concurrently. The fact that the hotter component is of slightly higher
flux than the cooler one (see Figure~\ref{specExamp}) provides clues
to the concentration of energy deposition in the persistent emission
activation zone. The increase of the RMS pulse fraction after the
sun-constrained down-time is possibly caused by the lower effective
emitting area of both components \citep[e.g.,][]{hu20ApJ:1818}.

The three distinct peaks in the profile at the onset of the
  outburst clearly indicate asymmetric, non-uniform heating pattern on
  the surface, yet with clear evolution towards a simpler, more
  localized region. In the magnetar model, it is natural to attribute
the soft X-ray active regions to being proximate to magnetic field
line footpoints on the surface. These footpoints may or may not be
associated with dipolar field morphology.  Perturbations to this
baseline field configuration have been developed in the {\sl twisted
  field scenario} \citep[e.g.,][]{thompson02ApJ:magnetars,
  2009ApJ...703.1044B}, via analogy to solar corona contexts. These
introduce toroidal field components driven by inner magnetospheric
currents, and it is possible that such magnetospheric field morphology
modifications could permeate magnetar atmospheres. Moreover the
currents will naturally bombard the stellar surface
\citep{BeloLi-2016-ApJ}, heating it at the footpoints of twisted field
lines.  Thus, the hot emission zones may be physically connected to
and/or co-located with bombardment footpoints. The pulse phase
migration and merging of the peaks seen in Figure~\ref{ppTimRes} as
the outburst evolves would suggest hot spot coalescence and field line
footpoint mobility.  The rate of pulse profile evolution can
potentially constrain twisted field models of magnetar activation,
suggesting insightful paths for future studies on the
theory/observation interface for magnetars.

Processes that originate in the crust may form an alternate scenario
to bombardment by twisted fields for the hot spot evolution. The rate
and character of evolution would be governed by bulk material
properties such as viscosity, heat conductivity, and the depth of
subsurface energy injection
\cite[e.g.,][]{2014ApJ...794L..24B,2016ApJ...824L..21L,2019MNRAS.486.4130L,2018Ap&SS.363..184K}. These
scenarios will be investigated in a separate study.

\subsection{Phase Distribution of Short Bursts}

Since our initial observation, \nicer\ frequently detected typical
short magnetar bursts from \src, revealing a total of 84 with an
average duration of about 30~ms. Much of the burst activity transpires
during the early stages of the outburst evolution, i.e., during the
steep flux decline and the strong pulse shape variation
(Figures~\ref{ppTimRes} and \ref{fig:spEv}. The fluence ranged from 3
to 30 $\times10^{-11}$~erg~cm$^{-2}$. Brighter bursts have also been
detected throughout the outburst by large field of view instruments
such as Fermi-GBM and Swift-BAT
\citep{fletcher21GCN1830,klingler21GCN1830}, though at a much lower
rate. This is qualitatively consistent with the steep $\log N$-$\log
S$ distribution of the \nicer\ events, $N(>F) \propto F^{-0.5}$. This
PL fluence distribution seems to be universal amongst magnetars
\citep[e.g.,][]{1996Natur.382..518C,gogus99apjl:1900,scholz11ApJ:1547,collazzi15ApJS,younes20ApJ1935},
and may be consistent with either the crust quake or magnetic
reconnection scenarios which are usually invoked as their triggering
mechanism
\citep[e.g.,][]{thompson95MNRAS:GF,thompson01ApJ:giantflare,2003MNRAS.346..540L,lyutikov15MNRAS,2016MNRAS.456.3282E}.

The apparent phase correlation of the bursts with that of the
persistent thermal surface hot-spot emission (Figure~\ref{fig:bursts})
is the strongest evidence yet of such phase-selection in observability
of short bursts in magnetars. Previously, a similar preference for
burst occurrence close to the persistent emission pulse maximum has
been reported in the radio-emitting magnetar XTE~J1810$-$197, however,
in a much smaller sample of four bursts
\citep{woods05ApJ:xte1810}. Observe that this phase variation of burst
arrival times contrasts the approximate phase-uniformity apparent in
the recent burst storm of SGR 1935+2154 \citep{younes20ApJ1935}.

This strong correlation suggests that approximate spatial colocation
of bursts and soft X-ray activity is highly likely, with the burst
plasma emitting at relatively low altitudes.  Given the $R^2 \sim 16$
km$^{2}$ apparent area of the bursts, one expects that occultation of
the bursts by the star at some pulse phases is readily achievable. We
envisage that the bursts could originate in twisted field line zones
near quasi-polar footpoints of closed field lines \citep{CB-2017-ApJ}.
Polarization-dependent radiative transport in optically-thick $e^+ \,
e^-$ plasmas in the presence of magnetar-strength B-fields could
impart strong beaming of radiation
\citep[e.g.,][]{2021MNRAS.500.5369B}, which could bias observability
of bursts at particular phases where the observer samples magnetic
field directions roughly parallel to the observer's line of sight.
This beaming is strong if the typical NICER-measured burst photon
energy of $3kT \sim 3-5$ keV is well below the local cyclotron energy
(roughly 2 MeV at the surface): such a domain arises if the altitude
of the bursts is less than about $5-7$ stellar radii assuming the
local magnetic field is not disrupted. These constraints pose a
challenge for reconnection models of magnetar bursts \citep[e.g.,
those invoking a relativistic tearing instability in an equatorial
current
sheet,][]{2003MNRAS.346..540L,2007MNRAS.374..415K,2016MNRAS.456.3282E}
and may point to a polar crustal origin
\citep[e.g.,][]{lander15mnras,2017ApJ...841...54T} for the mechanism
and locale of these magnetar short bursts. Future polarization studies
of magnetar bursts, for instance with LEAP \citep{2017HEAD...1610320M}
or AMEGO \citep{2019BAAS...51g.245M}, could elucidate the source
geometry and hone in on the altitude and magnetic colatitudes for
burst emission.

Finally, since magnetar short bursts could underpin some FRBs
\citep{2019ApJ...879....4W,2019MNRAS.488.5887S,Bochenek20:1935,
  chime2020:1935,2020ApJ...898L..29M}, our result of burst-phase
dependence has interesting implications for mechanisms, observability,
and periodic windowing of extragalactic FRBs, beyond the scope of this
paper.

\subsection{Radio behavior}

The non-detection of \src\ at MHz and GHz frequencies is commensurate
with the majority of magnetar sources. So far, only six magnetars have
been detected in the radio band. With the exception of SGR~1935+2154,
the properties of the radio emission from magnetars follow a general
trend; it is transient, usually appearing during outburst epochs, and
the radio spectrum is flat or inverted as opposed to the steep
spectrum detected in canonical radio pulsars
\citep[e.g.,][]{camilo06Natur:1810, torne15MNRAS:1745,
  pearlman2018ApJ,camilo16ApJ:1810, 2020ApJ...896L..30E}. Yet, the
radio emission from SGR~1935+2154 is unique to date. Hours after its
2020 April burst storm \citep{younes20ApJ1935}, the source emitted the
brightest radio flash ever detected from the Milky Way, with
properties reminiscent of extragalactic fast radio bursts
\citep{Bochenek20:1935, chime2020:1935},  simultaneous to a bright,
spectrally-unique short X-ray burst
\citep{mereghetti20ApJ,li21NatAs,younes2021NatAs, 2021NatAs...5..372R,
  2021NatAs...5..401T}. Many large radio campaigns ensued (weeks of
on-source observations), however, only three other radio bursts were
detected with two having fluences between 10 and 100 Jy ms
\citep{kirsten2020}. Our radio observations of \src\ were sensitive to
such brightness levels (Table~\ref{tab:radioObsSummary}). Furthermore,
months after SGR~1935+2154 had returned to quiescence, the FAST radio
telescope detected very weak pulsed emission from the source
\citep{zhu20ATel14084}. Such intermittent radio emission raises the
possibility that magnetars may be more prone to low-frequency
radiation than is currently thought. Hence, future regular radio
monitoring programs of magnetars both in outburst and during quiescent
periods may be key to unravelling the nature of the low-frequency
emission of these sources and their connection to extragalactic FRBs.

\section*{Acknowledgments}

G.Y. is partly funded through the NASA NICER GI program grant
80NSSC21K0233. M.G.B. acknowledges the generous support of the
National Science Foundation through grant
AST-1813649. Z.W. acknowledges support from the NASA postdoctoral
program. ZW acknowledges partial support by NASA under award number
80GSFC21M0002. F.K. acknowledges support from the Swedish Research
Council. A.B.P is a McGill Space Institute (MSI) Fellow and a Fonds de
Recherche du Quebec -- Nature et Technologies (FRQNT) postdoctoral
fellow. C.-P.H.~acknowledges support from the the Ministry of Science
and Technology in Taiwan through grant MOST
109-2112-M-018-009-MY3. This work is based in part on observations
carried out using the 32-m radio telescope operated by the Institute
of Astronomy of the Nicolaus Copernicus University in Toru\'n (Poland)
and supported by a Polish Ministry of Science and Higher Education
SpUB grant. A portion of this research was performed at the Jet
Propulsion Laboratory, California Institute of Technology and the
Caltech campus, under a Research and Technology Development Grant
through a contract with the National Aeronautics and Space
Administration. U.S. government sponsorship is
acknowledged. W.A.M. thanks the CDSCC staff and the DSN scheduling
team for their rapid response in scheduling and carrying out the radio
observations with the DSN.


\begin{thebibliography}{84}
\expandafter\ifx\csname natexlab\endcsname\relax\def\natexlab#1{#1}\fi

\bibitem[{{Agarwal} {et~al.}(2020){Agarwal}, {Aggarwal}, {Burke-Spolaor},
  {Lorimer}, \& {Garver-Daniels}}]{Agarwal2020MNRAS}
{Agarwal}, D., {Aggarwal}, K., {Burke-Spolaor}, S., {Lorimer}, D.~R., \&
  {Garver-Daniels}, N. 2020, \mnras, 497, 1661

\bibitem[{{Archibald} {et~al.}(2017){Archibald}, {Kaspi}, {Scholz},
  {Beardmore}, {Gehrels}, \& {Kennea}}]{archibald17ApJ:0142}
{Archibald}, R.~F., {Kaspi}, V.~M., {Scholz}, P., {et~al.} 2017, \apj, 834, 163

\bibitem[{{Archibald} {et~al.}(2016){Archibald}, {Kaspi}, {Tendulkar}, \&
  {Scholz}}]{archibald16:j1119}
{Archibald}, R.~F., {Kaspi}, V.~M., {Tendulkar}, S.~P., \& {Scholz}, P. 2016,
  \apjl, 829, L21

\bibitem[{{Archibald} {et~al.}(2020){Archibald}, {Scholz}, {Kaspi},
  {Tendulkar}, \& {Beardmore}}]{archibald20ApJ:1048}
{Archibald}, R.~F., {Scholz}, P., {Kaspi}, V.~M., {Tendulkar}, S.~P., \&
  {Beardmore}, A.~P. 2020, \apj, 889, 160

\bibitem[{{Barchas} {et~al.}(2021){Barchas}, {Hu}, \&
  {Baring}}]{2021MNRAS.500.5369B}
{Barchas}, J.~A., {Hu}, K., \& {Baring}, M.~G. 2021, \mnras, 500, 5369

\bibitem[Beloborodov(2009)]{2009ApJ...703.1044B} Beloborodov, A.~M.\
  2009, \apj, 703, 1044. doi:10.1088/0004-637X/703/1/1044

\bibitem[{{Beloborodov} \& {Levin}(2014)}]{2014ApJ...794L..24B}
{Beloborodov}, A.~M. \& {Levin}, Y. 2014, \apjl, 794, L24

\bibitem[{{Beloborodov} \& {Li}(2016)}]{BeloLi-2016-ApJ}
{Beloborodov}, A.~M. \& {Li}, X. 2016, \apj, 833, 261

\bibitem[{{Bochenek} {et~al.}(2020){Bochenek}, {Ravi}, {Belov}, {Hallinan},
  {Kocz}, {Kulkarni}, \& {McKenna}}]{Bochenek20:1935}
{Bochenek}, C.~D., {Ravi}, V., {Belov}, K.~V., {et~al.} 2020, \nat, 587, 59

\bibitem[{{Camilo} {et~al.}(2016){Camilo}, {Ransom}, {Halpern}, {Alford},
  {Cognard}, {Reynolds}, {Johnston}, {Sarkissian}, \& {van
  Straten}}]{camilo16ApJ:1810}
{Camilo}, F., {Ransom}, S.~M., {Halpern}, J.~P., {et~al.} 2016, \apj, 820, 110

\bibitem[{{Camilo} {et~al.}(2006){Camilo}, {Ransom}, {Halpern}, {Reynolds},
  {Helfand}, {Zimmerman}, \& {Sarkissian}}]{camilo06Natur:1810}
{Camilo}, F., {Ransom}, S.~M., {Halpern}, J.~P., {et~al.} 2006, \nat, 442, 892

\bibitem[{{Cash}(1979)}]{Cash1979}
{Cash}, W. 1979, \apj, 228, 939

\bibitem[{{Chen} \& {Beloborodov}(2017)}]{CB-2017-ApJ}
{Chen}, A.~Y. \& {Beloborodov}, A.~M. 2017, \apj, 844, 133

\bibitem[{{Cheng} {et~al.}(1996){Cheng}, {Epstein}, {Guyer}, \&
  {Young}}]{1996Natur.382..518C}
{Cheng}, B., {Epstein}, R.~I., {Guyer}, R.~A., \& {Young}, A.~C. 1996, \nat,
  382, 518

\bibitem[{{CHIME/FRB Collaboration} {et~al.}(2020){CHIME/FRB Collaboration},
  {Andersen}, {Bandura}, {Bhardwaj}, {Bij}, {Boyce}, {Boyle}, {Brar},
  {Cassanelli}, {Chawla}, {Chen}, {Cliche}, {Cook}, {Cubranic}, {Curtin},
  {Denman}, {Dobbs}, {Dong}, {Fandino}, {Fonseca}, {Gaensler}, {Giri}, {Good},
  {Halpern}, {Hill}, {Hinshaw}, {H{\"o}fer}, {Josephy}, {Kania}, {Kaspi},
  {Landecker}, {Leung}, {Li}, {Lin}, {Masui}, {McKinven}, {Mena-Parra},
  {Merryfield}, {Meyers}, {Michilli}, {Milutinovic}, {Mirhosseini},
  {M{\"u}nchmeyer}, {Naidu}, {Newburgh}, {Ng}, {Patel}, {Pen},
  {Pinsonneault-Marotte}, {Pleunis}, {Quine}, {Rafiei-Ravandi}, {Rahman},
  {Ransom}, {Renard}, {Sanghavi}, {Scholz}, {Shaw}, {Shin}, {Siegel}, {Singh},
  {Smegal}, {Smith}, {Stairs}, {Tan}, {Tendulkar}, {Tretyakov}, {Vanderlinde},
  {Wang}, {Wulf}, \& {Zwaniga}}]{chime2020:1935}
{CHIME/FRB Collaboration}, {Andersen}, B.~C., {Bandura}, K.~M., {et~al.} 2020,
  \nat, 587, 54

\bibitem[{{Collazzi} {et~al.}(2015){Collazzi}, {Kouveliotou}, {van der Horst},
  {Younes}, {Kaneko}, {G{\"o}{\u{g}}{\"u}{\textcommabelow s}}, {Lin}, {Granot},
  {Finger}, {Chaplin}, {Huppenkothen}, {Watts}, {von Kienlin}, {Baring},
  {Gruber}, {Bhat}, {Gibby}, {Gehrels}, {McEnery}, {van der Klis}, \&
  {Wijers}}]{collazzi15ApJS}
{Collazzi}, A.~C., {Kouveliotou}, C., {van der Horst}, A.~J., {et~al.} 2015,
  The Astrophysical Journal Supplement Series, 218, 11

\bibitem[{{Coti Zelati} {et~al.}(2021){Coti Zelati}, {Borghese}, {Israel},
  {Rea}, {Esposito}, {Pilia}, {Burgay}, {Possenti}, {Corongiu}, {Ridolfi},
  {Dehman}, {Vigan{\`o}}, {Turolla}, {Zane}, {Tiengo}, \&
  {Keane}}]{cotizelati21ApJ1830}
{Coti Zelati}, F., {Borghese}, A., {Israel}, G.~L., {et~al.} 2021, \apjl, 907,
  L34

\bibitem[{{Coti Zelati} {et~al.}(2018){Coti Zelati}, {Rea}, {Pons}, {Campana},
  \& {Esposito}}]{cotizelati18MNRAS}
{Coti Zelati}, F., {Rea}, N., {Pons}, J.~A., {Campana}, S., \& {Esposito}, P.
  2018, \mnras, 474, 961

\bibitem[{{Elenbaas} {et~al.}(2016){Elenbaas}, {Watts}, {Turolla}, \&
  {Heyl}}]{2016MNRAS.456.3282E}
{Elenbaas}, C., {Watts}, A.~L., {Turolla}, R., \& {Heyl}, J.~S. 2016, \mnras,
  456, 3282

\bibitem[{{Enoto} {et~al.}(2017){Enoto}, {Shibata}, {Kitaguchi}, {Suwa},
  {Uchide}, {Nishioka}, {Kisaka}, {Nakano}, {Murakami}, \&
  {Makishima}}]{2017ApJS..231....8E}
{Enoto}, T., {Shibata}, S., {Kitaguchi}, T., {et~al.} 2017, \apjs, 231, 8

\bibitem[Esposito et al.(2020)]{2020ApJ...896L..30E} Esposito, P.,
  Rea, N., Borghese, A., et al.\ 2020, \apjl, 896,
  L30. doi:10.3847/2041-8213/ab9742

\bibitem[{{Fletcher} \& {Fermi GBM Team}(2021)}]{fletcher21GCN1830}
{Fletcher}, C. \& {Fermi GBM Team}. 2021, GRB Coordinates Network, 29524, 1

\bibitem[{{Garmire} {et~al.}(2003){Garmire}, {Bautz}, {Ford}, {Nousek}, \&
  {Ricker}}]{garmire03SPIE}
{Garmire}, G.~P., {Bautz}, M.~W., {Ford}, P.~G., {Nousek}, J.~A., \& {Ricker},
  George~R., J. 2003, in Society of Photo-Optical Instrumentation Engineers
  (SPIE) Conference Series, Vol. 4851, X-Ray and Gamma-Ray Telescopes and
  Instruments for Astronomy., ed. J.~E. {Truemper} \& H.~D. {Tananbaum}, 28--44

\bibitem[Gavriil et al.(2008)]{gavriil2008Sci} Gavriil, F.~P.,
  Gonzalez, M.~E., Gotthelf, E.~V., et al.\ 2008, Science, 319,
  1802. doi:10.1126/science.1153465
    
\bibitem[{{Gavriil} {et~al.}(2011){Gavriil}, {Dib}, \&
  {Kaspi}}]{gavriil11ApJ:0142}
{Gavriil}, F.~P., {Dib}, R., \& {Kaspi}, V.~M. 2011, \apj, 736, 138

\bibitem[{{Gendreau} {et~al.}(2016){Gendreau}, {Arzoumanian}, {Adkins},
  {Albert}, {Anders}, {Aylward}, {Baker}, {Balsamo}, {Bamford}, {Benegalrao},
  {Berry}, {Bhalwani}, {Black}, {Blaurock}, {Bronke}, {Brown}, {Budinoff},
  {Cantwell}, {Cazeau}, {Chen}, {Clement}, {Colangelo}, {Coleman},
  {Coopersmith}, {Dehaven}, {Doty}, {Egan}, {Enoto}, {Fan}, {Ferro}, {Foster},
  {Galassi}, {Gallo}, {Green}, {Grosh}, {Ha}, {Hasouneh}, {Heefner}, {Hestnes},
  {Hoge}, {Jacobs}, {J{\o}rgensen}, {Kaiser}, {Kellogg}, {Kenyon}, {Koenecke},
  {Kozon}, {LaMarr}, {Lambertson}, {Larson}, {Lentine}, {Lewis}, {Lilly},
  {Liu}, {Malonis}, {Manthripragada}, {Markwardt}, {Matonak}, {Mcginnis},
  {Miller}, {Mitchell}, {Mitchell}, {Mohammed}, {Monroe}, {Montt de Garcia},
  {Mul{\'e}}, {Nagao}, {Ngo}, {Norris}, {Norwood}, {Novotka}, {Okajima},
  {Olsen}, {Onyeachu}, {Orosco}, {Peterson}, {Pevear}, {Pham}, {Pollard},
  {Pope}, {Powers}, {Powers}, {Price}, {Prigozhin}, {Ramirez}, {Reid},
  {Remillard}, {Rogstad}, {Rosecrans}, {Rowe}, {Sager}, {Sanders}, {Savadkin},
  {Saylor}, {Schaeffer}, {Schweiss}, {Semper}, {Serlemitsos}, {Shackelford},
  {Soong}, {Struebel}, {Vezie}, {Villasenor}, {Winternitz}, {Wofford},
  {Wright}, {Yang}, \& {Yu}}]{gendreau16SPIE}
{Gendreau}, K.~C., {Arzoumanian}, Z., {Adkins}, P.~W., {et~al.} 2016, Society
  of Photo-Optical Instrumentation Engineers (SPIE) Conference Series, Vol.
  9905, {The Neutron star Interior Composition Explorer (NICER): design and
  development}, 99051H

\bibitem[{{G{\"o}{\u g}{\"u}{\c s} } {et~al.}(1999){G{\"o}{\u g}{\"u}{\c s} },
  {Woods}, {Kouveliotou}, {van Paradijs}, {Briggs}, {Duncan}, \&
  {Thompson}}]{gogus99apjl:1900}
{G{\"o}{\u g}{\"u}{\c s} }, E., {Woods}, P.~M., {Kouveliotou}, C., {et~al.}
  1999, \apjl, 526, L93

\bibitem[{{G{\"o}{\u g}{\"u}{\c s}} {et~al.}(2020){G{\"o}{\u g}{\"u}{\c s}},
  {Kouveliotou}, \& {Younes}}]{gogus20ATel14085}
{G{\"o}{\u g}{\"u}{\c s}}, E., {Kouveliotou}, C., \& {Younes}, G. 2020, The
  Astronomer's Telegram, 14085, 1

\bibitem[{{G{\"o}{\u g}{\"u}{\c s}} {et~al.}(2016){G{\"o}{\u g}{\"u}{\c s}},
  {Lin}, {Kaneko}, {Kouveliotou}, {Watts}, {Chakraborty}, {Alpar},
  {Huppenkothen}, {Roberts}, {Younes}, \& {van der Horst}}]{gogus16:j1119}
{G{\"o}{\u g}{\"u}{\c s}}, E., {Lin}, L., {Kaneko}, Y., {et~al.} 2016, \apjl,
  829, L25

\bibitem[{{Hu} {et~al.}(2020){Hu}, {Begi{\c{c}}arslan}, {G{\"u}ver}, {Enoto},
  {Younes}, {Sakamoto}, {Ray}, {Strohmayer}, {Guillot}, {Arzoumanian},
  {Palmer}, {Gendreau}, {Malacaria}, {Wadiasingh}, {Jaisawal}, \&
  {Majid}}]{hu20ApJ:1818}
{Hu}, C.-P., {Begi{\c{c}}arslan}, B., {G{\"u}ver}, T., {et~al.} 2020, \apj,
  902, 1

\bibitem[{{Kaspi} \& {Beloborodov}(2017)}]{kaspi17:magnetars}
{Kaspi}, V.~M. \& {Beloborodov}, A. 2017, ArXiv e-prints

\bibitem[{{Kirsten} {et~al.}(2020){Kirsten}, {Snelders}, {Jenkins}, {Nimmo},
  {van den Eijnden}, {Hessels}, {Gawro{\'n}ski}, \& {Yang}}]{kirsten2020}
{Kirsten}, F., {Snelders}, M.~P., {Jenkins}, M., {et~al.} 2020, Nature
  Astronomy

\bibitem[{{Klingler} {et~al.}(2021){Klingler}, {Lien}, {Page}, \& {Neil Gehrels
  Swift Observatory Team}}]{klingler21GCN1830}
{Klingler}, N.~J., {Lien}, A.~Y., {Page}, K.~L., \& {Neil Gehrels Swift
  Observatory Team}. 2021, GRB Coordinates Network, 29516, 1

\bibitem[{{Komissarov} {et~al.}(2007){Komissarov}, {Barkov}, \&
  {Lyutikov}}]{2007MNRAS.374..415K}
{Komissarov}, S.~S., {Barkov}, M., \& {Lyutikov}, M. 2007, \mnras, 374, 415

\bibitem[{{Kouveliotou} {et~al.}(1998){Kouveliotou}, {Dieters}, {Strohmayer},
  {van Paradijs}, {Fishman}, {Meegan}, {Hurley}, {Kommers}, {Smith}, {Frail},
  \& {Murakami}}]{kouveliotou98Nat:1806}
{Kouveliotou}, C., {Dieters}, S., {Strohmayer}, T., {et~al.} 1998, \nat, 393,
  235

\bibitem[{{Kwang-Hua}(2018)}]{2018Ap&SS.363..184K}
{Kwang-Hua}, C.~W. 2018, \apss, 363, 184

\bibitem[{{Lander}(2016)}]{2016ApJ...824L..21L}
{Lander}, S.~K. 2016, \apjl, 824, L21

\bibitem[{{Lander} {et~al.}(2015){Lander}, {Andersson}, {Antonopoulou}, \&
  {Watts}}]{lander15mnras}
{Lander}, S.~K., {Andersson}, N., {Antonopoulou}, D., \& {Watts}, A.~L. 2015,
  \mnras, 449, 2047

\bibitem[{{Lander} \& {Gourgouliatos}(2019)}]{2019MNRAS.486.4130L}
{Lander}, S.~K. \& {Gourgouliatos}, K.~N. 2019, \mnras, 486, 4130

\bibitem[{{Leahy} \& {Tian}(2008)}]{leahy08AJ:w41}
{Leahy}, D.~A. \& {Tian}, W.~W. 2008, \aj, 135, 167

\bibitem[{{Li} {et~al.}(2021){Li}, {Lin}, {Xiong}, {Ge}, {Li}, {Li}, {Lu},
  {Zhang}, {Tuo}, {Nang}, {Zhang}, {Xiao}, {Chen}, {Song}, {Xu}, {Liu}, {Jia},
  {Cao}, {Qu}, {Zhang}, {Gu}, {Liao}, {Zhao}, {Tan}, {Nie}, {Zhao}, {Zheng},
  {Zheng}, {Luo}, {Cai}, {Li}, {Xue}, {Bu}, {Chang}, {Chen}, {Chen}, {Chen},
  {Chen}, {Chen}, {Cui}, {Cui}, {Deng}, {Dong}, {Du}, {Fu}, {Gao}, {Gao},
  {Gao}, {Gu}, {Guan}, {Guo}, {Han}, {Huang}, {Huo}, {Jiang}, {Jiang}, {Jin},
  {Jin}, {Kong}, {Li}, {Li}, {Li}, {Li}, {Li}, {Li}, {Li}, {Liang}, {Liu},
  {Liu}, {Liu}, {Liu}, {Liu}, {Lu}, {Lu}, {Luo}, {Ma}, {Meng}, {Ou}, {Sai},
  {Shang}, {Song}, {Sun}, {Tao}, {Wang}, {Wang}, {Wang}, {Wang}, {Wang}, {Wen},
  {Wu}, {Wu}, {Wu}, {Xiao}, {Xu}, {Yang}, {Yang}, {Yang}, {Yang}, {Yi}, {Yin},
  {You}, {Zhang}, {Zhang}, {Zhang}, {Zhang}, {Zhang}, {Zhang}, {Zhang},
  {Zhang}, {Zhang}, {Zhang}, {Zhang}, {Zhang}, {Zhang}, {Zhang}, {Zhang},
  {Zhang}, {Zhou}, {Zhou}, {Zhu}, {Zhu}, \& {Zhuang}}]{li21NatAs}
{Li}, C.~K., {Lin}, L., {Xiong}, S.~L., {et~al.} 2021, Nature Astronomy

\bibitem[Luo et al.(2021)]{luo19ascl:pint} Luo, J., Ransom, S.,
  Demorest, P., et al.\ 2021, \apj, 911,
  45. doi:10.3847/1538-4357/abe62f

\bibitem[{{Lyutikov}(2003)}]{2003MNRAS.346..540L}
{Lyutikov}, M. 2003, \mnras, 346, 540

\bibitem[{{Lyutikov}(2015)}]{lyutikov15MNRAS}
{Lyutikov}, M. 2015, \mnras, 447, 1407

\bibitem[{{Maan} {et~al.}(2020){Maan}, {Straal}, \& {van
  Leeuwen}}]{maan20ATel14098}
{Maan}, Y., {Straal}, S., \& {van Leeuwen}, J. 2020, The Astronomer's Telegram,
  14098, 1

\bibitem[{{Majid} {et~al.}(2017){Majid}, {Pearlman}, {Dobreva}, {Horiuchi},
  {Kocz}, {Lippuner}, \& {Prince}}]{Majid+2017}
{Majid}, W.~A., {Pearlman}, A.~B., {Dobreva}, T., {et~al.} 2017, \apjl, 834, L2

\bibitem[{{Majid} {et~al.}(2020){Majid}, {Pearlman}, {Nimmo}, {Hessels},
  {Prince}, {Naudet}, {Kocz}, \& {Horiuchi}}]{Majid+2020}
{Majid}, W.~A., {Pearlman}, A.~B., {Nimmo}, K., {et~al.} 2020, \apjl, 897, L4

\bibitem[{{McConnell} {et~al.}(2017){McConnell}, {Baring}, {Bloser}, {Briggs},
  {Connaughton}, {Dwyer}, {Gaskin}, {Grove}, {Gunji}, {Hartmann}, {Hayashida},
  {Hill}, {Kippen}, {Kishimoto}, {Kishimoto}, {Krizmanic}, {Lundman},
  {Mattingly}, {McBreen}, {Meegan}, {Mihara}, {Nakamori}, {Pearce}, {Phlips},
  {Preece}, {Produit}, {Ryan}, {Ryde}, {Sakamoto}, {Strickman}, {Sturner},
  {Takahashi}, {Toma}, {Vestrand}, {Wilson-Hodge}, {yatsu}, {Yonetoku}, \&
  {Zhang}}]{2017HEAD...1610320M}
{McConnell}, M.~L., {Baring}, M.~G., {Bloser}, P.~F., {et~al.} 2017, in
  AAS/High Energy Astrophysics Division, Vol.~16, AAS/High Energy Astrophysics
  Division \#16, 103.20

\bibitem[{{McEnery} {et~al.}(2019){McEnery}, {van der Horst}, {Dominguez},
  {Moiseev}, {Marcowith}, {Harding}, {Lien}, {Giuliani}, {Inglis}, {Ansoldi},
  {Stamerra}, {Manousakis}, {Strong}, {Bambi}, {Patricelli}, {Baring},
  {Barrio}, {Bastieri}, {Fields}, {Beacom}, {Beckmann}, {Bednarek}, {Rani},
  {Boggs}, {Bolotnikov}, {Cenko}, {Buckley}, {Grefenstette}, {Hui}, {Pittori},
  {Prescod-Weinstein}, {Shrader}, {Gouiffes}, {Kierans}, {Wilson-Hodge},
  {D'Ammando}, {Castro}, {Kocveski}, {Gasparrini}, {Thompson}, {Williams}, {De
  Angelis}, {Bernard}, {Digel}, {Morcuende}, {Charles}, {Bissaldi}, {Hays},
  {Ferrara}, {Bozzo}, {Grove}, {Wulf}, {Bottacini}, {Caroli}, {Kislat},
  {Oikonomou}, {Giordano}, {Longo}, {Fryer}, {Fukazawa}, {Georganopoulos}, {De
  Nolfo}, {Vianello}, {Kanbach}, {Younes}, {Blumer}, {Hartmann}, {Hernanz},
  {Takahashi}, {Li}, {Agudo}, {Moskalenko}, {Stumke}, {Grenier}, {Smith},
  {Rodi}, {Perkins}, {Gelfand}, {Holder}, {Knodlseder}, {Kopp}, {Lenain},
  {{\'A}lvarez}, {Metcalfe}, {Krizmanic}, {Stephen}, {Hewitt}, {Mitchell},
  {Harding}, {Tomsick}, {Racusin}, {Finke}, {Kargaltsev}, {Klimenko},
  {Krawczynski}, {Smith}, {Kubo}, {Di Venere}, {Marcotulli}, {Lommler},
  {Parker}, {Baldini}, {Foffano}, {Zampieri}, {Tibaldo}, {Petropoulou},
  {Ajello}, {Meyer}, {L{\'o}pez}, {McConnell}, {Boettcher}, {Cardillo},
  {Martinez}, {Kerr}, {Mazziotta}, {McEnery}, {Di Mauro}, {Wood}, {Meyer},
  {Briggs}, {De Becker}, {Lovellette}, {Doro}, {Sanchez-Conde}, {Moss},
  {Mizuno}, {Rib{\'o}}, {Nakazawa}, {Neilson}, {Auricchio}, {Omodei},
  {Oberlack}, {Ohno}, {Orlando}, {Otte}, {Coppi}, {Bloser}, {Zhang}, {Laurent},
  {Pohl}, {Prandini}, {Shawhan}, {Caputo}, {Campana}, {Rando}, {Woolf},
  {Johnson}, {Mignani}, {Walter}, {Ojha}, {da Silva}, {Dietrich}, {Funk},
  {Zane}, {Anton}, {Buson}, {Cutini}, {Saz Parkinson}, {Schirato}, {Griffin},
  {Kaufmann}, {Stawarz}, {Ciprini}, {Del Sordo}, {Jones}, {Guiriec}, {Tajima},
  {Cheung}, {The}, {Venters}, {Porter}, {Linden}, {Barres}, {Paliya},
  {Bozhilov}, {Vestrand}, {Tatischeff}, {Chen}, {Wang}, {Tanaka}, {Uhm},
  {Zhang}, {Zimmer}, {Zoglauer}, \& {Wadiasingh}}]{2019BAAS...51g.245M}
{McEnery}, J., {van der Horst}, A., {Dominguez}, A., {et~al.} 2019, in Bulletin
  of the American Astronomical Society, Vol.~51, 245

\bibitem[{{Mereghetti}(2008)}]{mereghetti08AARv:magentars}
{Mereghetti}, S. 2008, \aapr, 15, 225

\bibitem[{{Mereghetti} {et~al.}(2020{\natexlab{a}}){Mereghetti}, {Savchenko},
  {Ferrigno}, {G{\"o}tz}, {Rigoselli}, {Tiengo}, {Bazzano}, {Bozzo}, {Coleiro},
  {Courvoisier}, {Doyle}, {Goldwurm}, {Hanlon}, {Jourdain}, {von Kienlin},
  {Lutovinov}, {Martin-Carrillo}, {Molkov}, {Natalucci}, {Onori}, {Panessa},
  {Rodi}, {Rodriguez}, {S{\'a}nchez-Fern{\'a}ndez}, {Sunyaev}, \&
  {Ubertini}}]{2020ApJ...898L..29M}
{Mereghetti}, S., {Savchenko}, V., {Ferrigno}, C., {et~al.} 2020{\natexlab{a}},
  \apjl, 898, L29

\bibitem[{{Mereghetti} {et~al.}(2020{\natexlab{b}}){Mereghetti}, {Savchenko},
  {Ferrigno}, {G{\"o}tz}, {Rigoselli}, {Tiengo}, {Bazzano}, {Bozzo}, {Coleiro},
  {Courvoisier}, {Doyle}, {Goldwurm}, {Hanlon}, {Jourdain}, {von Kienlin},
  {Lutovinov}, {Martin-Carrillo}, {Molkov}, {Natalucci}, {Onori}, {Panessa},
  {Rodi}, {Rodriguez}, {S{\'a}nchez-Fern{\'a}ndez}, {Sunyaev}, \&
  {Ubertini}}]{mereghetti20ApJ}
{Mereghetti}, S., {Savchenko}, V., {Ferrigno}, C., {et~al.} 2020{\natexlab{b}},
  \apjl, 898, L29

\bibitem[{{Page} {et~al.}(2020){Page}, {Barthelmy}, {Klingler}, {Kuin}, \&
  {Lien}}]{page20ATel14083}
{Page}, K.~L., {Barthelmy}, S.~D., {Klingler}, N.~J., {Kuin}, N.~P.~M., \&
  {Lien}, A.~Y. 2020, The Astronomer's Telegram, 14083, 1

\bibitem[{{Pearlman} {et~al.}(2019){Pearlman}, {Majid}, \&
  {Prince}}]{pearlman+2019a}
{Pearlman}, A.~B., {Majid}, W.~A., \& {Prince}, T.~A. 2019, Advances in
  Astronomy, 2019, 6325183

\bibitem[{{Pearlman} {et~al.}(2020{\natexlab{a}}){Pearlman}, {Majid}, {Prince},
  {Bansal}, {Horiuchi}, {Stephenson}, {Naudet}, \& {Kocz}}]{pearlman+2020a}
{Pearlman}, A.~B., {Majid}, W.~A., {Prince}, T.~A., {et~al.}
  2020{\natexlab{a}}, The Astronomer's Telegram, 14102, 1

\bibitem[{{Pearlman} {et~al.}(2018{\natexlab{a}}){Pearlman}, {Majid}, {Prince},
  {Kocz}, \& {Horiuchi}}]{Pearlman+2018}
{Pearlman}, A.~B., {Majid}, W.~A., {Prince}, T.~A., {Kocz}, J., \& {Horiuchi},
  S. 2018{\natexlab{a}}, \apj, 866, 160

\bibitem[{{Pearlman} {et~al.}(2018{\natexlab{b}}){Pearlman}, {Majid}, {Prince},
  {Kocz}, \& {Horiuchi}}]{pearlman2018ApJ}
{Pearlman}, A.~B., {Majid}, W.~A., {Prince}, T.~A., {Kocz}, J., \& {Horiuchi},
  S. 2018{\natexlab{b}}, \apj, 866, 160

\bibitem[{{Pearlman} {et~al.}(2020{\natexlab{b}}){Pearlman}, {Majid}, {Prince},
  {Nimmo}, {Hessels}, {Naudet}, \& {Kocz}}]{Pearlman+2020b}
{Pearlman}, A.~B., {Majid}, W.~A., {Prince}, T.~A., {et~al.}
  2020{\natexlab{b}}, arXiv e-prints, arXiv:2009.13559

\bibitem[{{Pearlman} {et~al.}(2020{\natexlab{c}}){Pearlman}, {Majid}, {Prince},
  {Ray}, {Kocz}, {Horiuchi}, {Naudet}, {G{\"u}ver}, {Enoto}, {Arzoumanian},
  {Gendreau}, \& {Ho}}]{Pearlman+2020c}
{Pearlman}, A.~B., {Majid}, W.~A., {Prince}, T.~A., {et~al.}
  2020{\natexlab{c}}, arXiv e-prints, arXiv:2005.08410

\bibitem[{{Ray} {et~al.}(2011){Ray}, {Kerr}, {Parent}, {Abdo}, {Guillemot},
  {Ransom}, {Rea}, {Wolff}, {Makeev}, {Roberts}, {Camilo}, {Dormody}, {Freire},
  {Grove}, {Gwon}, {Harding}, {Johnston}, {Keith}, {Kramer}, {Michelson},
  {Romani}, {Saz Parkinson}, {Thompson}, {Weltevrede}, {Wood}, \&
  {Ziegler}}]{ray11ApJS}
{Ray}, P.~S., {Kerr}, M., {Parent}, D., {et~al.} 2011, \apjs, 194, 17

\bibitem[{{Ray} {et~al.}(2020){Ray}, {Younes}, {Guver}, {Wadiasingh},
  {Arzoumanian}, {Gendreau}, {Enoto}, {Ho}, {Hu}, {Bansal}, \&
  {Majid}}]{ray20ATel14112}
{Ray}, P.~S., {Younes}, G., {Guver}, T., {et~al.} 2020, The Astronomer's
  Telegram, 14112, 1

\bibitem[{{Rea} {et~al.}(2016){Rea}, {Borghese}, {Esposito}, {Coti Zelati},
  {Bachetti}, {Israel}, \& {De Luca}}]{rea16:rcw103}
{Rea}, N., {Borghese}, A., {Esposito}, P., {et~al.} 2016, \apjl, 828, L13

\bibitem[{{Rea} {et~al.}(2013){Rea}, {Israel}, {Pons}, {Turolla}, {Vigan{\`o}},
  {Zane}, {Esposito}, {Perna}, {Papitto}, {Terreran}, {Tiengo}, {Salvetti},
  {Girart}, {Palau}, {Possenti}, {Burgay}, {G{\"o}{\u g}{\"u}{\c s}},
  {Caliandro}, {Kouveliotou}, {G{\"o}tz}, {Mignani}, {Ratti}, \&
  {Stella}}]{rea13ApJ:0418}
{Rea}, N., {Israel}, G.~L., {Pons}, J.~A., {et~al.} 2013, \apj, 770, 65

\bibitem[{{Remillard} {et~al.}(2021){Remillard}, {Loewenstein}, {Steiner},
  {Prigozhin}, {LaMarr}, {Enoto}, {Gendreau}, {Arzoumanian}, {Markwardt},
  {Basak}, {Stevens}, {Ray}, {Altamirano}, \& {Buisson}}]{remillard2021:3c50}
{Remillard}, R.~A., {Loewenstein}, M., {Steiner}, J.~F., {et~al.} 2021, arXiv
  e-prints, arXiv:2105.09901

\bibitem[Ridnaia et al.(2021)]{2021NatAs...5..372R} Ridnaia, A.,
  Svinkin, D., Frederiks, D., et al.\ 2021, Nature Astronomy, 5,
  372. doi:10.1038/s41550-020-01265-0

\bibitem[{{Scargle} {et~al.}(2013){Scargle}, {Norris}, {Jackson}, \&
  {Chiang}}]{scargle2013apj:BB}
{Scargle}, J.~D., {Norris}, J.~P., {Jackson}, B., \& {Chiang}, J. 2013, \apj,
  764, 167

\bibitem[{{Scholz} \& {Kaspi}(2011)}]{scholz11ApJ:1547}
{Scholz}, P. \& {Kaspi}, V.~M. 2011, \apj, 739, 94

\bibitem[{{Surnis} {et~al.}(2020){Surnis}, {Joshi}, {Stappers}, {Sand},
  {Bagchi}, {Rajwade}, {Jankowski}, {Caleb}, \& {Gajjar}}]{surnis20ATel14091}
{Surnis}, M., {Joshi}, B.~C., {Stappers}, B., {et~al.} 2020, The Astronomer's
  Telegram, 14091, 1

\bibitem[{{Suvorov} \& {Kokkotas}(2019)}]{2019MNRAS.488.5887S}
{Suvorov}, A.~G. \& {Kokkotas}, K.~D. 2019, \mnras, 488, 5887

\bibitem[Tavani et al.(2021)]{2021NatAs...5..401T} Tavani, M.,
  Casentini, C., Ursi, A., et al.\ 2021, Nature Astronomy, 5,
  401. doi:10.1038/s41550-020-01276-x

\bibitem[{{Thompson} \& {Duncan}(1995)}]{thompson95MNRAS:GF}
{Thompson}, C. \& {Duncan}, R.~C. 1995, \mnras, 275, 255

\bibitem[{{Thompson} \& {Duncan}(1996)}]{thompson96ApJ:magnetar}
{Thompson}, C. \& {Duncan}, R.~C. 1996, \apj, 473, 322

\bibitem[{{Thompson} \& {Duncan}(2001)}]{thompson01ApJ:giantflare}
{Thompson}, C. \& {Duncan}, R.~C. 2001, \apj, 561, 980

\bibitem[{{Thompson} {et~al.}(2002){Thompson}, {Lyutikov}, \&
  {Kulkarni}}]{thompson02ApJ:magnetars}
{Thompson}, C., {Lyutikov}, M., \& {Kulkarni}, S.~R. 2002, \apj, 574, 332

\bibitem[{{Thompson} {et~al.}(2017){Thompson}, {Yang}, \&
  {Ortiz}}]{2017ApJ...841...54T}
{Thompson}, C., {Yang}, H., \& {Ortiz}, N. 2017, \apj, 841, 54

\bibitem[{{Tian} \& {Leahy}(2008)}]{tian08ApJkes73}
{Tian}, W.~W. \& {Leahy}, D.~A. 2008, \apj, 677, 292

\bibitem[{{Torne} {et~al.}(2015){Torne}, {Eatough}, {Karuppusamy}, {Kramer},
  {Paubert}, {Klein}, {Desvignes}, {Champion}, {Wiesemeyer}, {Kramer},
  {Spitler}, {Thum}, {G{\"u}sten}, {Schuster}, \&
  {Cognard}}]{torne15MNRAS:1745}
{Torne}, P., {Eatough}, R.~P., {Karuppusamy}, R., {et~al.} 2015, \mnras, 451,
  L50

\bibitem[{{Turolla} {et~al.}(2015){Turolla}, {Zane}, \&
  {Watts}}]{turolla15:mag}
{Turolla}, R., {Zane}, S., \& {Watts}, A.~L. 2015, Reports on Progress in
  Physics, 78, 116901

\bibitem[{{Vigan{\`o}} {et~al.}(2013){Vigan{\`o}}, {Rea}, {Pons}, {Perna},
  {Aguilera}, \& {Miralles}}]{vigano13MNRAS}
{Vigan{\`o}}, D., {Rea}, N., {Pons}, J.~A., {et~al.} 2013, Monthly Notices of
  the Royal Astronomical Society, 434, 123

\bibitem[{{Wadiasingh} \& {Timokhin}(2019)}]{2019ApJ...879....4W}
{Wadiasingh}, Z. \& {Timokhin}, A. 2019, \apj, 879, 4

\bibitem[{{Whitney} {et~al.}(2010){Whitney}, {Kettenis}, {Phillips}, \&
  {Sekido}}]{2010ivs..conf..192W}
{Whitney}, A., {Kettenis}, M., {Phillips}, C., \& {Sekido}, M. 2010, in Sixth
  International VLBI Service for Geodesy and Astronomy. Proceedings from the
  2010 General Meeting, ed. R.~{Navarro}, S.~{Rogstad}, C.~E. {Goodhart},
  E.~{Sigman}, M.~{Soriano}, D.~{Wang}, L.~A. {White}, \& C.~S. {Jacobs},
  192--196

\bibitem[{{Woods} {et~al.}(2004){Woods}, {Kaspi}, {Thompson}, {Gavriil},
  {Marshall}, {Chakrabarty}, {Flanagan}, {Heyl}, \&
  {Hernquist}}]{woods04ApJ:1E2259}
{Woods}, P.~M., {Kaspi}, V.~M., {Thompson}, C., {et~al.} 2004, \apj, 605, 378

\bibitem[{{Woods} {et~al.}(2005){Woods}, {Kouveliotou}, {Gavriil}, {Kaspi},
  {Roberts}, {Ibrahim}, {Markwardt}, {Swank}, \& {Finger}}]{woods05ApJ:xte1810}
{Woods}, P.~M., {Kouveliotou}, C., {Gavriil}, F.~P., {et~al.} 2005, \apj, 629,
  985

\bibitem[{{Younes} {et~al.}(2021){Younes}, {Baring}, {Kouveliotou},
  {Arzoumanian}, {Enoto}, {Doty}, {Gendreau}, {G{\"o}{\v{g}}{\"u}{\c{s}}},
  {Guillot}, {G{\"u}ver}, {Harding}, {Ho}, {van der Horst}, {Hu}, {Jaisawal},
  {Kaneko}, {LaMarr}, {Lin}, {Majid}, {Okajima}, {Pope}, {Ray}, {Roberts},
  {Saylor}, {Steiner}, \& {Wadiasingh}}]{younes2021NatAs}
{Younes}, G., {Baring}, M.~G., {Kouveliotou}, C., {et~al.} 2021, Nature
  Astronomy

\bibitem[{{Younes} {et~al.}(2020{\natexlab{a}}){Younes}, {G{\"u}ver},
  {Kouveliotou}, {Baring}, {Hu}, {Wadiasingh}, {Begi{\c{c}}arslan}, {Enoto},
  {G{\"o}{\u{g}}{\"u}{\c{s}}}, {Lin}, {Harding}, {van der Horst}, {Majid},
  {Guillot}, \& {Malacaria}}]{younes20ApJ1935}
{Younes}, G., {G{\"u}ver}, T., {Kouveliotou}, C., {et~al.} 2020{\natexlab{a}},
  \apjl, 904, L21

\bibitem[{{Younes} {et~al.}(2020{\natexlab{b}}){Younes}, {Guver}, {Wadiasingh},
  {Gendreau}, {Arzoumanian}, {Ray}, {Ho}, {Majid}, \&
  {Enoto}}]{younes20ATel14086}
{Younes}, G., {Guver}, T., {Wadiasingh}, Z., {et~al.} 2020{\natexlab{b}}, The
  Astronomer's Telegram, 14086, 1

\bibitem[{{Younes} {et~al.}(2016){Younes}, {Kouveliotou}, {Kargaltsev}, {Gill},
  {Granot}, {Watts}, {Gelfand}, {Baring}, {Harding}, {Pavlov}, {van der Horst},
  {Huppenkothen}, {G{\"o}{\u g}{\"u}{\c s}}, {Lin}, \&
  {Roberts}}]{younes16ApJ:1834}
{Younes}, G., {Kouveliotou}, C., {Kargaltsev}, O., {et~al.} 2016, \apj, 824,
  138

\bibitem[{{Younes} {et~al.}(2015){Younes}, {Kouveliotou}, \&
  {Kaspi}}]{younes15ApJ:1806}
{Younes}, G., {Kouveliotou}, C., \& {Kaspi}, V.~M. 2015, \apj, 809, 165

\bibitem[{{Zhu} {et~al.}(2020){Zhu}, {Wang}, {Zhou}, {Xu}, {Wang}, {Zhang},
  {Feng}, {Han}, {Jiang}, {Lee}, {Di}, {Lin}, {Men}, {Niu}, {Xu}, {Yang}, {Yu},
  {Zhang}, {Zhang}, {Zhang}, {Zhang}, \& {Zhang}}]{zhu20ATel14084}
{Zhu}, W., {Wang}, B., {Zhou}, D., {et~al.} 2020, The Astronomer's Telegram,
  14084, 1

\end{thebibliography}
\end{document}

We report on NICER X-ray monitoring of the magnetar SGR~1830-0645 covering 223 days following its October 2020 outburst, as well as Chandra and radio observations. We present the most accurate spin ephemerides of the source so far: f=0.096~Hz, f_dot=-6.20e-14~Hz~s-1, and a significant second and third frequency derivative terms indicative of non-negligible timing noise. The phase-averaged 0.8-7~keV spectrum is well fit with 2 blackbody models throughout the campaign. The BB temperatures remain constant at 0.46 and 1.2~keV. The areas and flux of each component shrank by a factor of 6, initially through a steep decay trend lasting about 46 days followed by a shallow long-term one. The pulse shape in the same energy range is initially complex, exhibiting three distinct peaks, yet with clear continuous evolution throughout the outburst towards a simpler, single-pulse shape. The rms pulsed fraction is high and increases from about 40\% to 50\%. We find no dependence of pulse shape or fraction on energy. These results suggest that multiple hotspots, possibly possessing temperature gradients, emerged at outburst-onset, and shrank as the outburst decayed. We detect 84 faint bursts with NICER, having a strong preference for occurring close to the surface emission pulse maximum; the first time this phenomenon is detected in such a large burst sample. This likely implies a very low altitude for the burst emission region, and a triggering mechanism connected to the surface active zone. Finally, our radio observations at several epochs and multiple frequencies reveal no evidence of pulsed or burst-like radio emission.